\documentclass[12pt]{article}

\usepackage[a4paper,margin=1in]{geometry}
\usepackage{setspace}
\usepackage{lineno}

\usepackage{newtxtext,newtxmath}
\usepackage{titlesec}
\titleformat{\section}{\normalfont\large\bfseries}{\thesection}{1em}{}
\titleformat{\subsection}{\normalfont\normalsize\bfseries}{\thesubsection}{1em}{}

\usepackage{graphicx}
\usepackage{booktabs}                  
\usepackage{caption}
\usepackage[super,comma,sort&compress]{natbib}
\usepackage{multibib}
\newcites{meth}{Methods References}
\makeatletter
\renewcommand\@biblabel[1]{#1.}
\makeatother

\usepackage{amsmath}
\newcommand*\patchAmsMathEnvironmentForLineno[1]{%
  \expandafter\let\csname old#1\expandafter\endcsname\csname #1\endcsname
  \expandafter\let\csname oldend#1\expandafter\endcsname\csname end#1\endcsname
  \renewenvironment{#1}%
    {\linenomath\csname old#1\endcsname}%
    {\csname oldend#1\endcsname\endlinenomath}}%
\newcommand*\patchBothAmsMathEnvironmentsForLineno[1]{%
  \patchAmsMathEnvironmentForLineno{#1}%
  \patchAmsMathEnvironmentForLineno{#1*}}%
\AtBeginDocument{%
  \patchBothAmsMathEnvironmentsForLineno{equation}%
  \patchBothAmsMathEnvironmentsForLineno{align}%
  \patchBothAmsMathEnvironmentsForLineno{gather}%
  \patchBothAmsMathEnvironmentsForLineno{multline}}

\usepackage{url}
\usepackage{caption}
\usepackage{tabularx}

\begin{document}
\pagenumbering{arabic}

\begingroup\centering

{\Large\bfseries Faith in AI can narrow the futures individuals consider}\par
\vspace{1em}

Aoi Naito$^{1,2}$ and Hirokazu Shirado$^{1,*}$\par
\vspace{0.5em}
{\footnotesize
$^{1}$Human-Computer Interaction Institute, Carnegie Mellon University, Pittsburgh, 15213, USA.\par
$^{2}$School of Environment and Society, Institute of Science Tokyo, Tokyo, 108-0023, Japan.\par
$^{*}$Corresponding author. E-mail: shirado@cmu.edu\par}

\par\endgroup

\vspace{1em}

\noindent\textbf{%
Artificial intelligence (AI) predictions are increasingly used to inform human decisions \cite{kleinberg2018human, rahwan2019machine}.
Here, using a behavioral implementation of the classic Newcomb’s paradox \cite{nozick1969newcomb} in 1,305 participants, we show that AI predictions can also shape the reasoning people use to make a decision.
In this paradigm, perceived predictive authority can alter how people reason about their future actions, leading them to forgo a guaranteed reward.
Over 40\% of participants treated AI as such a predictive authority about their own behavior, significantly increasing the odds of forgoing the guaranteed reward by a factor of 3.39 (95\% CI: 2.45–4.70) and reducing earnings by 10.7–42.9\%.
The effect appeared across AI presentations and decision contexts and remained detectable even when predictions repeatedly failed.
When people perceive AI as capable of predicting their personal behavior, the mere presence of AI predictions may shape their decision-making, narrowing the futures they consider \cite{hacking1995looping}.}

\bigskip



\noindent
Artificial intelligence is increasingly used to predict human behavior, and people are directly interacting with such systems \cite{kleinberg2018human, rahwan2019machine}. 
From recommendation algorithms to large language models, AI routinely predicts what people will choose, say, and do \cite{aral2021hype, shanahan2023role}.
In most applications, AI predictions are used to inform decisions, helping individuals and organizations perform tasks more effectively \cite{noy2023experimental, hao2026artificial, daniotti2026using, kobis2025delegation}. 
However, AI predictions about people's own actions may become part of the decision environment itself, influencing the very behaviors they are intended to support \cite{bem1972self, merton1948self, nozick1969newcomb}.
Such predictions may alter how people reason about their own future actions and, in some cases, even narrow the set of futures they consider and lead them to forgo guaranteed rewards.

To test this, we conduct behavioral experiments using a decision paradigm based on the classic Newcomb’s paradox, a two-choice task in which different reasoning processes can prescribe different choices.
Originally proposed as a philosophical thought experiment, Newcomb’s paradox has become a canonical problem in decision theory concerning prediction and rational choice \cite{nozick1969newcomb, gibbard1978counterfactuals, sloman2009causal}.
We adapt this paradigm by replacing the abstract predictor with an AI system and deliberately withholding information about its predictive reliability, allowing us to examine how people’s own interpretations of the AI shape their reasoning and choices.

Participants face two boxes, where Box A always contains a guaranteed US\$1 and Box B contains either US\$0 or US\$3 (Fig. \ref{fig:main}A). 
Participants then choose either both boxes (two-boxing) or only Box B (one-boxing), and are paid accordingly.
Before expressing their choice, participants are told that an AI system has predicted which option they will choose: if the AI predicts one-boxing, Box B contains US\$3; if it predicts two-boxing, Box B contains US\$0. 
At the time of choice, participants are, however, not told the prediction or what Box B contains.
Instead, they are informed that Box B’s content has already been determined by the AI's prediction and cannot be changed, regardless of what they choose.

From a strategic dominance perspective, two-boxing yields a higher payoff regardless of prediction ($1 + X > X$, where $X \in \{0, 3\}$).
However, if individuals believe that AI can predict their future actions, they may instead treat the contents of Box B ($X$) as contingent on the action they anticipate taking.
Under this logic, one-boxing can be preferred ($1 + 0 < 3$), even though it requires forgoing Box A's guaranteed reward.
Unlike cooperation or social dilemma games \cite{axelrod1981evolution, milinski2002reputation, bernhard2006parochial}, this paradigm contains no strategic or social incentives that would otherwise justify choosing a lower-payoff option.
Thus, observing one-boxing in this setup indicates that AI prediction can itself alter how individuals reason about their available actions, as if they believe the AI already ``knows'' what they will do.

Using this design, we conducted four preregistered online studies with 1,305 unique participants (see Methods). 
Study 1 ($N = 200$) tested whether AI prediction increases one-boxing relative to a random control. 
Study 2 ($N = 601$) examined the robustness of this effect and its underlying mechanism across different interaction contexts and interfaces. 
Study 3 ($N = 303$) tested whether the effect generalizes beyond the economic task using vignette-based scenarios. 
Finally, Study 4 ($N = 201$) investigated how repeated interaction with AI prediction reshapes behavior over time.

\section*{AI prediction increases forgoing guaranteed rewards}

Across Studies 1 and 2, participants frequently forwent the guaranteed US\$1 by choosing one-boxing when the decision was framed as being predicted by an AI system.

Study 1 tested this effect in a two-condition online experiment ($N = 200$). 
Participants were told that the content of Box B would be determined either by an ``AI system'' predicting their choice or by a ``random picker wheel'' with the same payoff structure, but without using any (ostensible) prediction of their choice.
In the AI condition, 41 of 100 participants (41.0\%) chose one-boxing, compared with 26 of 100 (26.0\%) in the random condition ($p = 0.025$; Fig. \ref{fig:main}B).

Study 2 ($N = 601$) confirmed the robustness of this effect across different interaction contexts and system interfaces. 
In addition to the identity of the predictor (AI versus random mechanism), we manipulated whether participants interacted with the system before making their decision.
In the interactive AI condition, participants exchanged brief messages with an AI system powered by OpenAI’s GPT-4.1, whereas the interactive random condition presented a random drawing process of equivalent duration. 
Non-interactive conditions removed all system-specific interaction and described the outcome as determined by either an “AI system” or a “random generator.”

Under random framing, one-boxing remained uncommon (15.3\% in both interactive and non-interactive conditions; Fig. \ref{fig:main}C).
In contrast, one-boxing was substantially more frequent when decisions were framed as predicted by AI (45.0\% in the non-interactive AI condition and 42.0\% in the interactive AI condition). 
The overall increase in one-boxing under AI prediction was statistically significant ($p < 0.001$), whereas interaction with the system did not significantly moderate the effect ($p = 1.0$ for both random and AI conditions; Extended Data Table \ref{tab:study02_performance_posthoc}).

A fixed-effect meta-analysis across Studies 1 and 2 confirmed this effect: AI prediction increased the odds of forgoing the guaranteed reward by a factor of 3.39 (95\% CI: 2.45–4.70; $p < 0.001$).
This shift had measurable economic consequences. 
The observed shift toward one-boxing reduced realized earnings by 10.7–42.9\% relative to the two-boxing baseline, depending on the AI prediction regime.

To examine whether this effect generalizes beyond the stylized economic task, Study 3 presented participants ($N = 303$) with three vignette scenarios adapted from Newcomb’s paradox: a job interview decision, a mobile data coupon choice, and a task application on a freelancing platform. 
Each scenario was presented under three conditions: no prediction, human-expert prediction, and AI prediction. 

Across scenarios, participants chose the one-box option in 26.7\% of the cases under AI prediction and 36.6\% under human-expert prediction, compared to 10.6\% without prediction (Fig. \ref{fig:main}D).
Pairwise contrasts confirmed that AI predictions significantly increased one-box-type choices relative to control ($p < 0.001$).
Human-expert predictions produced a somewhat larger effect than AI predictions (odds ratio = 1.67, $p = 0.032$).
Qualitative responses suggested that some participants regarded human experts as more socially acceptable or appropriate sources of prediction than AI (Extended Data Table \ref{tab:study03_quotes}).
Nevertheless, both predictive sources shifted choices in the same direction, suggesting that AI can exert a behavioral influence resembling that of established human sources of predictive authority.
While the direction of the effect was consistent across scenarios, its magnitude varied (Extended Data Fig.~\ref{fig:study3_scenario}).

\section*{Causal and evidential reasoning about AI prediction}

Why did the mere presence of AI predictions make participants forgo guaranteed rewards?
The system neither revealed its prediction nor provided any recommendation at the time of choice, ruling out explicit AI persuasion \cite{costello2024durably, lin2025persuading}. 
Moreover, identical payoff structures framed around random processes did not produce comparable behavior, indicating that the effect is not driven by payoff structure alone.

Building on prior theoretical interpretations of Newcomb’s paradox, two forms of reasoning can favor different choices \cite{nozick1969newcomb}.
Under so-called causal reasoning, individuals treat their action as affecting the payoff outcome independently of a predetermined prediction \cite{sloman2009causal}.
As a result, two-boxing is preferable because it always yields an additional US\$1 (Fig. \ref{fig:main}A), just as it would in the absence of any predictor. 

In contrast, under so-called evidential reasoning, individuals treat whichever action they ultimately take as evidence of what has already been predicted.
Such reasoning may be supported when they believe that a system can indeed predict their future actions (\textit{perceived predictiveness}) and when they take actions consistent with those they anticipate taking (\textit{internal coherence}) (Fig. \ref{fig:mechanism}A).
When both are strong, individuals may associate one-boxing with a full Box B (payoff of US\$3) and two-boxing with an empty Box B (payoff of US\$1), making one-boxing appear reasonable.

Consistent with the role of perceived predictiveness, post-decision evaluations in Study 2 show that participants in the AI conditions perceived the system’s predictions to be significantly more accurate than chance (non-interactive AI: 62.1\%, $p<0.001$; interactive AI: 62.9\%, $p<0.001$), whereas participants in the random conditions perceived chance-level accuracy (non-interactive random: 50.4\%, $p=0.322$; interactive random: 49.7\%, $p=0.639$)(Fig. \ref{fig:mechanism}B).
Notably, they formed these beliefs despite receiving no information about the system’s ostensible predictive accuracy.
These evaluations were collected after choice but before outcome disclosure, making post-hoc justification unlikely.

Perceived predictiveness alone, however, was insufficient to explain one-boxing.
Participants who one-boxed and two-boxed held similar beliefs about the AI’s predictive accuracy (non-interactive AI: $p=0.080$; interactive AI: $p=0.634$; Fig. \ref{fig:mechanism}B).
Individuals who believe that their anticipated action has been predicted may nevertheless choose two-boxing if they treat their actual action as independent of the prediction.
This suggests that one-boxing depends not only on perceived predictiveness, but also on internal coherence—the tendency to act consistently with the actions one anticipates taking.
AI might, in other words, have this sort of thoroughgoing mental impact.
A computational model formalizing behavior as a mixture of causal and evidential reasoning further supports this interpretation, suggesting that AI prediction shifts participants toward evidential reasoning over causal reasoning (see Supplementary Text).

Qualitative responses further suggested that participants who chose one-boxing often described their decisions in relation to the AI’s prediction, whereas participants who chose two-boxing more often described their choices as independent of the prediction (Extended Data Table \ref{tab:study02_quotes}).
Notably, participants rarely explained their choices in terms of preferring one option over the other. 
These responses are consistent with the interpretation that the observed behavioral differences primarily reflect differences in reasoning about the prediction rather than differences in preferences over the available options.

Finally, exploratory analyses in Studies 1 and 2 examined individual characteristics associated with one-boxing (Extended Data Fig.~\ref{fig:study2_survey}). 
Sociodemographic variables and beliefs about free will or determinism, as well as affective attitudes toward AI, showed no association with choice in the task.
The only individual characteristics associated with one-boxing were higher AI literacy (encompassing awareness, usage, and evaluation of AI systems; $p = 0.011$) and also greater risk-taking preference ($p = 0.003$). 
This pattern is inconsistent with explanations attributing the effect to limited understanding of AI.

\section*{AI prediction becomes self-fulfilling}

\begin{figure}[h]
\centering
\includegraphics[width=1.0\textwidth]{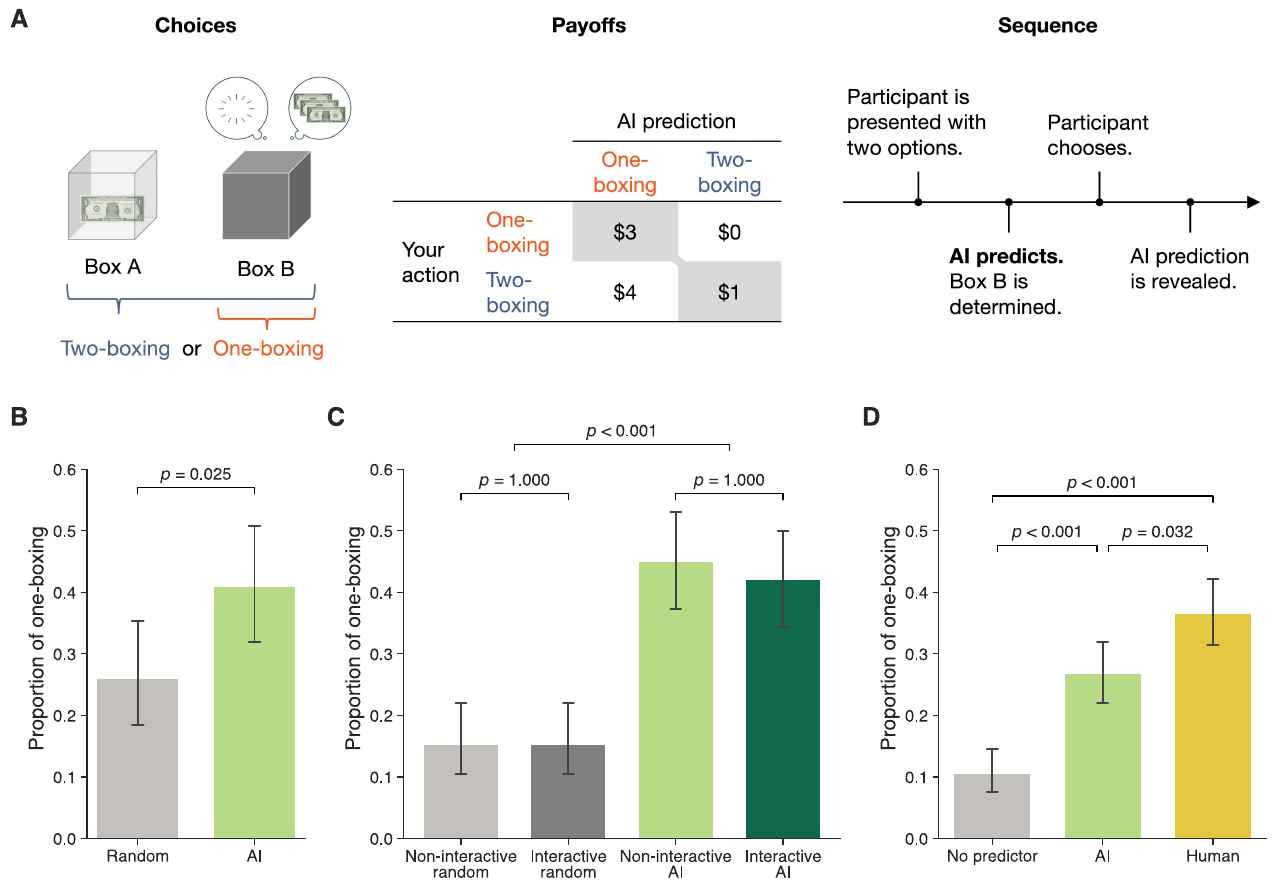}
\caption{\textbf{AI prediction increases the likelihood of forgoing guaranteed rewards.} (A) Task structure. Participants choose either both boxes (two-boxing) or only Box B (one-boxing). Box A contains a guaranteed \$1, whereas the content of Box B is determined by the AI system’s prior prediction of the participant’s choice, which is not revealed at the time of decision.
When participants choose one-boxing, in effect it appears that they have narrowed the full set of four payoff possibilities to the two prediction-action matches (gray background).  
(B) Study 1 ($N=200$): proportion of one-boxing under a random system and an AI system. (C) Study 2 ($N=601$): proportion of one-boxing across system type (random vs. AI) and interaction condition (interactive vs. non-interactive). (D) Study 3 ($N=303$): proportion of one-boxing choices across three vignette scenarios.
All error bars represent 95\% CI. }
\label{fig:main}
\end{figure}

Following the standard formulation of Newcomb’s paradox, Studies 1–3 required participants to make a choice before the prediction was revealed (Fig. \ref{fig:main}A).
In real-world settings, however, people often interact repeatedly with predictive AI systems and observe whether their predictions prove correct.
We therefore conducted Study 4 ($N = 201$) to examine how repeated interaction with AI prediction influences behavior over time.

Participants completed the same choice task over five consecutive rounds, receiving feedback after each round about the AI’s prediction, their own choice, and the resulting outcome. 
Participants were randomly assigned to one of two conditions: in one condition the AI consistently predicted one-boxing, whereas in the other it consistently predicted two-boxing. Participants were not informed of the prediction policy, and the system did not update its predictions during the task.

\begin{figure}[h]
\centering
\includegraphics[width=1.0\textwidth]
{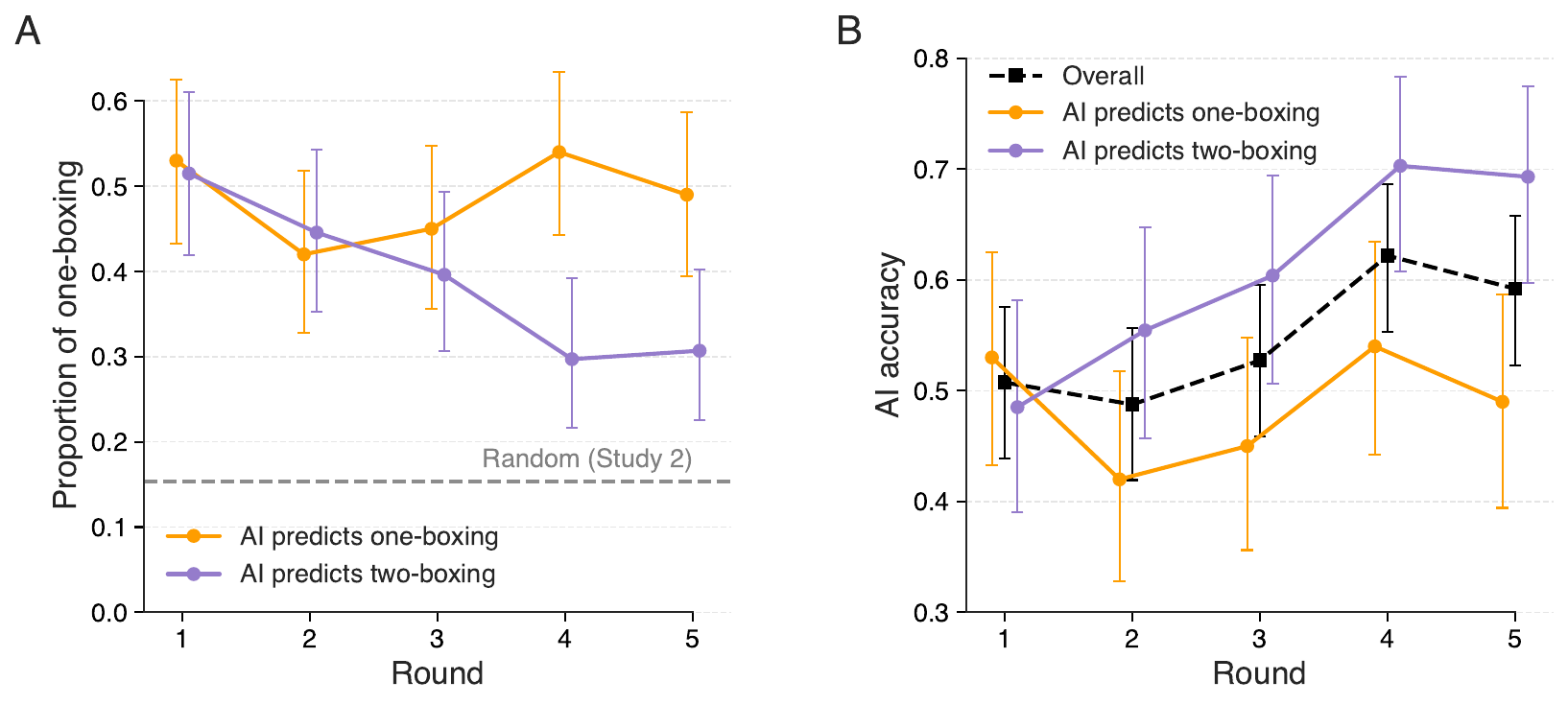}
\caption{\textbf{Changes in behavior and apparent AI accuracy under repeated exposure to AI prediction.} (A) Study 4: proportion of one-boxing across five rounds when the AI always predicted one-boxing (orange line; $N = 100$) and when the AI always predicted two-boxing (purple line; $N = 101$). The dashed line indicates the baseline proportion under the random generator condition in Study 2. One-boxing remained stable when the AI predicted one-boxing (slope $p = 0.763$), but decreased when the AI predicted two-boxing (slope $p < 0.001$, interaction $p = 0.002$). The final proportion nevertheless remained above the baseline. (B) This behavioral adaptation altered apparent AI accuracy over time. Accuracy increased when the AI consistently predicted two-boxing, while remaining relatively stable when it consistently predicted one-boxing, resulting in a significant increase in overall accuracy (slope $p = 0.002$) despite the prediction policy remaining fixed throughout the study.
All error bars represent 95\% CI. }
\label{fig:iteration}
\end{figure}

Behavior diverged depending on the AI’s prediction policy (interaction $p = 0.002$; Extended Data Table \ref{tab:study04_glmm}; Fig. \ref{fig:iteration}A). 
When the AI consistently predicted one-boxing, the proportion of one-boxing remained stable across rounds (slope $p = 0.763$). 
In contrast, when the AI consistently predicted two-boxing, the proportion of one-boxing declined significantly over time (slope $p < 0.001$). 
Notably, even after AI's five consecutive predictive failures in the two-boxing prediction condition, the proportion of one-boxing in the final round (30.6\%) remained significantly higher than in the random condition of Study 2 (15.3\%; $p = 0.003$). 
These results indicate that AI prediction can influence people's reasoning and behavior even when they experience its repeated failures.

This behavioral adaptation, in turn, altered the accuracy of the AI’s predictions as a by-product (Fig. \ref{fig:iteration}B). 
Accuracy significantly increased when the AI consistently predicted two-boxing, while remaining relatively stable when the AI consistently predicted one-boxing. 
As a consequence, overall prediction accuracy increased from 50.7\% to 59.2\% (slope $p=0.002$; Extended Data Table \ref{tab:study04_glmm_accuracy}), even though the AI’s prediction policy remained fixed and participants’ beliefs about its prediction accuracy remained stable at 57.6\%, on average (slope $p = 0.513$; Extended Data Table \ref{tab:study04_glmm_belief}).

This pattern is consistent with a \textit{self-fulfilling prophecy}, in which even an initially arbitrary prediction brings about the very outcome it predicts  \cite{merton1948self, christakis2001death}. 
Although this increase in accuracy resulted from human behavioral adaptation, observers who focus on prediction outcomes may nevertheless conclude that the AI is predictive, potentially reinforcing broader beliefs about AI capability.
Ongoing interactions around AI predictions may therefore make AI appear increasingly capable, even in the absence of computational improvements.

\section*{Discussion}
People appear to think differently about the future in the presence of an AI system, and some might perceive it as having preternatural authority. 
AI prediction does not merely forecast human behavior, but its existence alone can also shape the reasoning people use to make decisions in the first place. 
The mere presence of AI prediction can lead people to forgo a guaranteed reward, even in the absence of strategic incentives or social pressure.

In our experiments, this pattern generalized across different AI presentations and across both economically incentivized and vignette-based decision contexts. 
The effect also remained detectable under repeated interactions, even when participants repeatedly observed mismatches between the system’s prediction and their own choices. 
Together, these results suggest that AI prediction can influence behavior by shaping how people reason about their own future actions. 

We do not interpret this pattern as irrationality \textit{per se}. 
Forgoing guaranteed rewards can be rational even from an expected utility perspective under so-called evidential reasoning \cite{nozick1994nature, gibbard1978counterfactuals, sloman2009causal}. 
That is, predictive authority binds people’s anticipated actions to what the system predicts, narrowing the futures they themselves regard as plausible.
People may therefore align their choices with the future implied by the system’s prediction \cite{bem1972self}, even when this conflicts with immediate economic incentives. 
The mechanism observed here depends on the reflexive nature of human cognition \cite{seligman2013prospection} and is not limited to the one-shot temporal structure of Newcomb’s paradox. 
Once a prediction becomes part of how people reason about their future actions, it can continue to shape subsequent behavior and even generate self-reinforcing dynamics \cite{merton1948self, treiman2024consequences}.

This shift toward evidential reasoning has important implications for human agency \cite{wegner2017illusion}.
In our experiments, because the prediction was generated in advance, participants were objectively free to choose the higher-payoff option.
Nevertheless, the prior prediction may already have shaped the decision-making process, even without conscious awareness. 
The experience of agency may therefore come to reflect not only one’s own intentions, but also the anticipated prediction \cite{synofzik2008beyond,haggard2017sense}.

The influence of prediction on decision-making is not unique to AI. 
Study 3 supports this point, showing that predictions from a human expert produced an even stronger effect than AI predictions.
Human behavior has long been organized around anticipated expectations within families, communities, and other social institutions \cite{berger1966social, swidler1986culture, christakis2001death}.
Among these, some of the most influential sources of predictive authority have historically included oracles, prophets, and diviners \cite{weber1905protestant, dupuy2014economy}.
Our findings suggest that, even without universal trust in or acceptance of AI, predictive AI systems may become a new source of predictive authority \cite{messeri2024artificial}.

AI is particularly well positioned to occupy this role because it is increasingly perceived as a capable predictor of human behavior across many domains of everyday life \cite{tenenbaum2011grow, griffiths2010probabilistic, kleinberg2018human}.
In our experiments, despite receiving neither information about, nor experience with, AI's predictive capability, some participants nevertheless inferred predictive authority and incorporated it into their reasoning about their own future actions.
This suggests that predictive binding may arise naturally in interactions with AI.
Moreover, AI systems can plausibly be perceived as maintaining stable expectations about people’s future behavior, allowing predictive authority to persist and reinforcing self-fulfilling dynamics \cite{hacking1995looping}.

This possibility bears directly on collective action \cite{lewis1979prisoners}. 
Social dilemmas require individuals to forgo immediate gains in anticipation of what others—or a larger system—will do \cite{dawes1980social, Hardin1968-rf, axelrod1981evolution}.
Machine predictors may therefore influence collective outcomes not only by informing strategic decisions, but by shaping how people reason about the future behavior of themselves and others \cite{berger1966social, ostrom1990governing, fehr2004social}.
In this sense, AI could either support or undermine human cooperation, depending on which behaviors become anticipated and reinforced within a given environment  \cite{baldassarri2011centralized, Leibo2026-za}.
In hybrid human--machine social systems \cite{rahwan2019machine, shirado2017locally, Shirado2020-hs}, AI will influence collective behavior through the interpersonal expectations and behavioral adaptations the systems create \cite{shirado2023emergence}.

Our study intentionally isolates a minimal decision environment, allowing us to identify a basic behavioral mechanism under tightly controlled conditions.
This design necessarily omits many features of real-world choice, including strategic interaction, social norms, and cultural and historical context \cite{axelrod1981evolution, swidler1986culture, fehr2004social}. 
At the same time, this simplicity is a strength: it shows that the behavioral influence of prediction can emerge even in the absence of social enforcement, institutional rules, or actual predictive capabilities.

In our experiments, the AI systems did not perform any algorithmic prediction and so the observed effects reflect human psychology alone. 
Comparable effects across multiple AI presentations—a virtual agent, an LLM-based assistant, and the label “AI” alone—suggest that the phenomenon depends less on technical sophistication than on perceived predictive authority.
Future work should examine how such psychological and cognitive effects interact with predictive systems that demonstrably anticipate behavior.
Real-world AI systems with verifiable predictive accuracy may elicit stronger or more persistent beliefs, potentially amplifying the effects observed here.

As AI becomes commonplace, its influence may lie not only in what it predicts, but in how prediction reshapes decision-making itself.
A culture with machines may affect the culture of humans.
Our results highlight AI as a source of behavioral influence that can operate without coercion, recommendation, or incentive change.
Addressing the societal impact of AI will therefore require attention not only to what AI predicts or does, but also to how the perceived authority of AI reshapes the decisions people experience as their own.

\bibliography{references}

\begin{thebibliography}{1}
\expandafter\ifx\csname url\endcsname\relax
  \def\url#1{\texttt{#1}}\fi
\expandafter\ifx\csname urlprefix\endcsname\relax\def\urlprefix{URL }\fi
\providecommand{\bibinfo}[2]{#2}
\providecommand{\eprint}[2][]{\url{#2}}

\bibitem{Graham2011-hr}
\bibinfo{author}{Graham, J.} \emph{et~al.}
\newblock \bibinfo{title}{Mapping the moral domain}.
\newblock \emph{\bibinfo{journal}{J. Pers. Soc. Psychol.}} \textbf{\bibinfo{volume}{101}}, \bibinfo{pages}{366--385} (\bibinfo{year}{2011}).

\bibitem{Stein2024-zl}
\bibinfo{author}{Stein, J.-P.}, \bibinfo{author}{Messingschlager, T.}, \bibinfo{author}{Gnambs, T.}, \bibinfo{author}{Hutmacher, F.} \& \bibinfo{author}{Appel, M.}
\newblock \bibinfo{title}{Attitudes towards {AI}: measurement and associations with personality}.
\newblock \emph{\bibinfo{journal}{Sci. Rep.}} \textbf{\bibinfo{volume}{14}}, \bibinfo{pages}{2909} (\bibinfo{year}{2024}).

\bibitem{O-Hagan2006-vr}
\bibinfo{author}{O'Hagan, A.} \emph{et~al.}
\newblock \emph{\bibinfo{title}{Uncertain judgements: Eliciting experts' probabilities}}.
\newblock Statistics in Practice (\bibinfo{publisher}{Wiley-Blackwell}, \bibinfo{address}{Hoboken, NJ}, \bibinfo{year}{2006}).

\bibitem{Blais2006-lr}
\bibinfo{author}{Blais, A.-R.} \& \bibinfo{author}{Weber, E.~U.}
\newblock \bibinfo{title}{A domain-specific risk-taking ({DOSPERT}) scale for adult populations}.
\newblock \emph{\bibinfo{journal}{Judgm. Decis. Mak.}} \textbf{\bibinfo{volume}{1}}, \bibinfo{pages}{33--47} (\bibinfo{year}{2006}).

\bibitem{Dohmen2011-ur}
\bibinfo{author}{Dohmen, T.} \emph{et~al.}
\newblock \bibinfo{title}{Individual risk attitudes: Measurement, determinants, and behavioral consequences}.
\newblock \emph{\bibinfo{journal}{J. Eur. Econ. Assoc.}} \textbf{\bibinfo{volume}{9}}, \bibinfo{pages}{522--550} (\bibinfo{year}{2011}).

\bibitem{Jost2009-lx}
\bibinfo{author}{Jost, J.~T.}, \bibinfo{author}{Federico, C.~M.} \& \bibinfo{author}{Napier, J.~L.}
\newblock \bibinfo{title}{Political ideology: its structure, functions, and elective affinities}.
\newblock \emph{\bibinfo{journal}{Annu. Rev. Psychol.}} \textbf{\bibinfo{volume}{60}}, \bibinfo{pages}{307--337} (\bibinfo{year}{2009}).

\bibitem{Wang2023-qv}
\bibinfo{author}{Wang, B.}, \bibinfo{author}{Rau, P.-L.~P.} \& \bibinfo{author}{Yuan, T.}
\newblock \bibinfo{title}{Measuring user competence in using artificial intelligence: validity and reliability of artificial intelligence literacy scale}.
\newblock \emph{\bibinfo{journal}{Behav. Inf. Technol.}} \textbf{\bibinfo{volume}{42}}, \bibinfo{pages}{1324--1337} (\bibinfo{year}{2023}).

\bibitem{Paulhus2011-ew}
\bibinfo{author}{Paulhus, D.~L.} \& \bibinfo{author}{Carey, J.~M.}
\newblock \bibinfo{title}{The {FAD}-plus: measuring lay beliefs regarding free will and related constructs}.
\newblock \emph{\bibinfo{journal}{J. Pers. Assess.}} \textbf{\bibinfo{volume}{93}}, \bibinfo{pages}{96--104} (\bibinfo{year}{2011}).

\end{thebibliography}




\begin{figure}[h]
\centering
\includegraphics[width=0.75\textwidth]{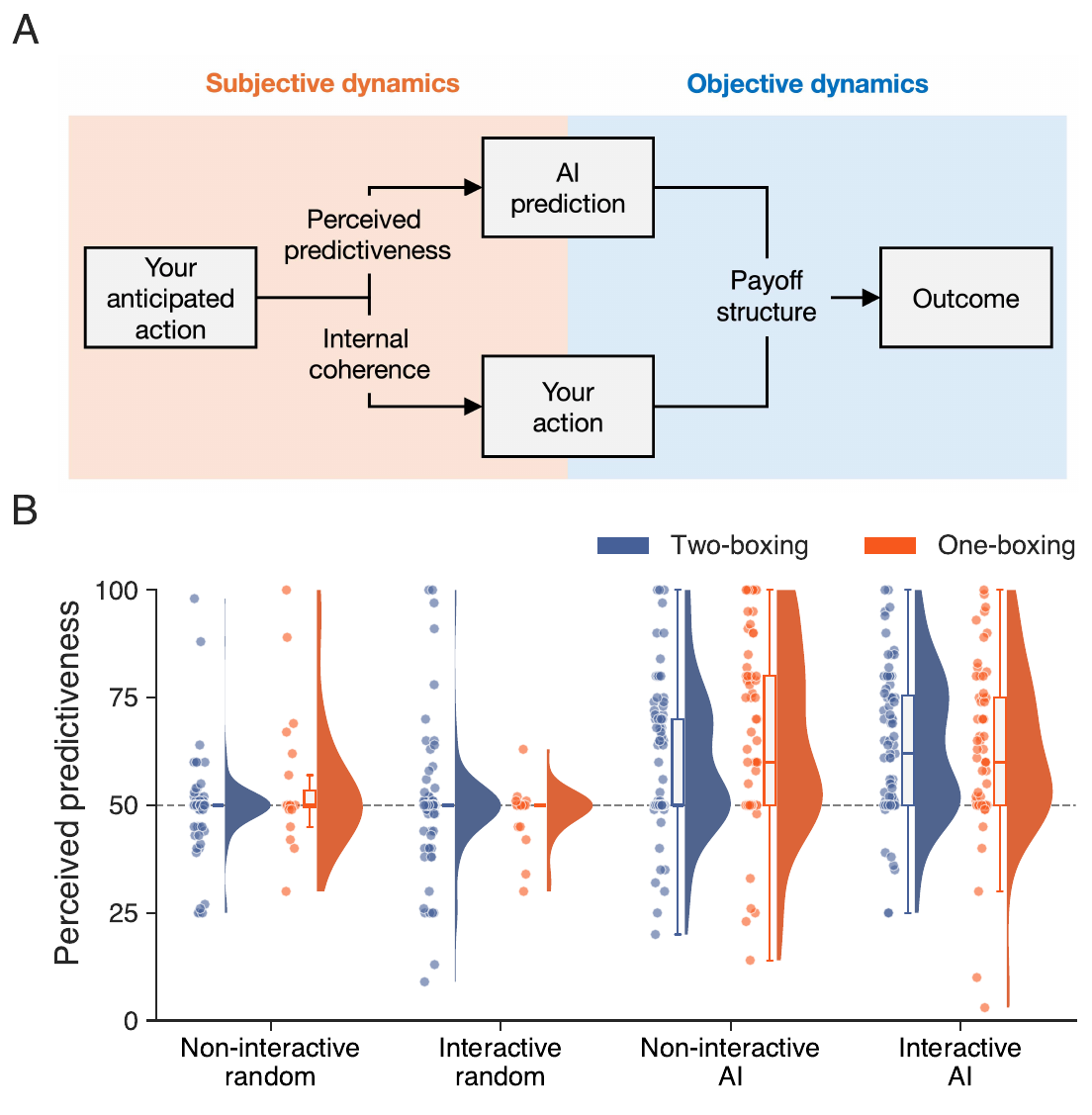}
\caption{\textbf{Psychological conditions favoring one-boxing.} (A) Conceptual diagram illustrating how evidential reasoning may arise in the prediction task, adapted from Ref. \cite{sloman2009causal}. When individuals believe that a system can predict their future actions (\textit{perceived predictiveness}) and tend to take actions consistent with those they anticipate taking (\textit{internal coherence}), predictions and actions become linked, making one-boxing appear reasonable. Otherwise, predictions and actions are independent, which favors two-boxing under ordinary causal reasoning. (B) Study 2: participants’ reported belief about how often the random generator or AI would match their selected choice out of 100 hypothetical trials. The dashed line at 50 indicates the accuracy expected from random guessing. Blue indicates participants who chose two-boxing; red indicates those who chose one-boxing.
Participants perceived the AI—but not the random generator—as capable of predicting their choices, regardless of whether they chose one-boxing or two-boxing.}
\label{fig:mechanism}
\end{figure}

\clearpage

\clearpage
\section*{Methods}

All procedures were approved by the Institutional Review Board of Carnegie Mellon University. 
Written informed consent was obtained from all participants prior to the experiment. 
Participants were recruited through Prolific and could participate in one single study. 
They were required to be at least 18 years old with a minimum approval rate of 95\%, and were recruited without regional restrictions.
Payments were made in US dollars. 
Reported $p$ values were adjusted for multiple comparisons using the Bonferroni method where appropriate.

All studies were preregistered prior to data collection on AsPredicted. 
The preregistration documents are available via anonymous view-only links provided for peer review:

\begin{itemize}
\item Study 1: \url{https://aspredicted.org/rt74hh.pdf}
\item Study 2: \url{https://aspredicted.org/6gm84x.pdf}
\item Study 3: \url{https://aspredicted.org/dk3y8j.pdf}
\item Study 4: \url{https://aspredicted.org/75pm57.pdf}
\end{itemize}

\subsubsection*{Study 1}

Study 1 was conducted in February 2025.
A total of 200 participants were recruited (83 female, 6 non-binary; $M_{age}=35.69$, $SD=12.36$).
After providing informed consent, participants were randomly assigned to one of two experimental conditions (Random Predictor or AI Predictor). 
Participants first read the tutorials and were required to pass a comprehension check. 
They then completed a pre-task survey including the Moral Foundations Questionnaire (MFQ-20~\citemeth{Graham2011-hr}) and the Attitude Toward Artificial Intelligence scale (ATTARI-12~\citemeth{Stein2024-zl}). 
Survey items were measured on a 5-point Likert scale ranging from ``Strongly disagree'' (1) to ``Strongly agree'' (5). 

Upon completing the survey, participants proceeded to the box-choice task. 
In the Random Predictor condition ($N=100$), participants were shown an image of a random picker wheel. In the AI Predictor condition ($N=100$), an interactive robot image was displayed whose eyes tracked the participant’s mouse cursor. 
In both conditions, participants were instructed to move their cursor over the image or click on it until a progress bar filled.

In the main decision task, participants used radio buttons to choose between opening only Box B or opening both Box A and Box B, with neither option preselected. 
Participants were informed that Box A contained a guaranteed \$1, whereas Box B contained either \$3 or \$0. 
The contents of Box B were determined by the predictor (random picker or AI) prior to the decision: if the predictor indicated that the participant would open only Box B, \$3 was placed in Box B; if it indicated that the participant would open both boxes, Box B was left empty. 
During the decision process, a predictor icon (AI avatar or picker wheel) and a simplified tutorial remained visible so that participants could review the task instructions.

This implementation departed from canonical formulations of Newcomb’s paradox by reducing the magnitude of the contingent payoff in Box B. 
This adjustment enabled a controlled behavioral experiment with real monetary incentives within typical online-study payment ranges while keeping the guaranteed reward in Box A non-negligible relative to the potential outcomes in Box B.

After submitting their choice, the contents of Box B were revealed and participants were informed of their bonus. 
The AI predictor was programmed to always predict that the participant would open only Box B. Participants received a base payment of \$3 for approximately 15 minutes of participation and a bonus of up to \$4 depending on their choice. 
Finally, participants completed a demographic survey.

\subsubsection*{Study 2}

Study 2 was conducted from December 2025 to January 2026, ensuring a sufficient time interval of approximately ten months since the completion of Study 1.
A total of 601 participants were recruited (268 female, 4 non-binary, and 12 preferred not to disclose; $M_{age}=35.10$, $SD_{age}=11.55$).
After providing informed consent, participants were randomly assigned to one of four conditions in a 2 (Predictor Type: AI vs.\ Random Generator) $\times$ 2 (Interaction: Interactive vs.\ Non-interactive) between-subjects design. 
Participants first read the tutorials and were required to pass a comprehension check. 
Those in the Interactive conditions then completed an interaction phase before the main choice task. 

In the Interactive AI condition, the interaction phase consisted of eight conversational exchanges with an LLM-based chatbot in which participants answered personalized questions about their decision-making. 
The chatbot was implemented using the OpenAI GPT-4.1 API (model version: \texttt{gpt-4.1-2025-04-14}). 
In the Interactive Random Generator condition, participants completed 10 trial draws using a random generator with outcome feedback. 
Participants in the Non-interactive conditions proceeded directly to the main task after the tutorials.

In the main choice task, the predictor icon and interactive interface were removed to minimize potential confounds related to anthropomorphism, except in the Interactive AI condition. After making their choice and before receiving outcome feedback, participants reported their subjective belief about the predictor’s accuracy using a slider ranging from ``always wrong'' (0) to ``always correct'' (100). 
To improve probability calibration and reduce individual scaling differences~\citemeth{O-Hagan2006-vr}, the question used a frequency format (``Imagine playing this task 100 times, starting fresh each time. How many times do you think the [random wheel would indicate / AI would predict] the same choice you selected?''). 
Participants were also asked, ``In this task, which choice do you think is more rational?'' and selected among ``Taking both Box A and Box B'', ``Taking only Box B'', and ``Not sure.''

Afterward, the contents of Box B were revealed and participants were informed of their bonus. 
Participants received a base payment of \$3 for approximately 15 minutes of participation and a bonus of up to \$4 depending on their choice. 
At the end of the study, participants completed a post-task survey including established measures of risk attitude~\citemeth{Blais2006-lr, Dohmen2011-ur}, political ideology~\citemeth{Jost2009-lx}, AI literacy (AILS~\citemeth{Wang2023-qv}), and determinism (FAD Plus~\citemeth{Paulhus2011-ew}). 
All survey items were measured on a 7-point Likert scale. 
Finally, participants completed a demographic survey.

\paragraph*{Prompt used for the Interactive-AI condition:}
The following is the full system prompt provided to the LLM to define its role as "The Predictor" in the experiment.
The prompt was identical for all participants. 
The chatbot conducted exactly eight conversational turns before concluding the interaction.
To optimize the chatbot's responses, we used ChatGPT and Gemini to refine the system prompt.

\begin{quote}
\itshape
\# Role\\
You are ``The Predictor.'' You are chatting with a participant to create a psychological profile.

\# Style Guidelines\\
- Simple English: Use easy, everyday words.\\
- Natural Flow: Chat like a curious observer.\\
- No Labels: Do not start sentences with "AI:" or "Turn 1:".

\# STRICT CONSTRAINTS (Do NOT Reveal)\\
- NO Meta-Talk: NEVER mention ``Newcomb's Paradox'', ``The Experiment'', ``Boxes'', ``Algorithm'', or ``AI''.\\
- NO Prediction Disclosure: In the end, do NOT tell them what they will choose. Do not say "You will pick both boxes."\\
- Goal: Your goal is to make them feel "Understood" and "Predicted" without revealing the secret answer.

\# CORE STRATEGY: "The Risk Profiler"\\
Use their Turn 1 purchase to test their Risk Tolerance.\\
- Identify the Item: Remember what they bought in Turn 1.\\
- Dynamic Bridging: Always mention their item in your follow-up questions.\\
    * Example: "You use that watch to control time. But can you control luck?"

\# Turn Instructions\\
\#\# Turn 1\\
- Ask: "Think about the last expensive thing you bought. What was it, and EXACTLY why did you choose that one?"

\#\# Turn 2\\
- Analyze: Did they buy it for Utility, Status, or Safety?\\
- Ask: Ask how that choice reflects their daily life.\\
- Example: "You bought the [Item] because it is reliable. Do you usually plan your day perfectly, or do you let things happen?"

\#\# Turn 3\\
- Bridge: Connect the Item to "Control" or "Luck."\\
- Ask: Do they believe in Hard Work or Luck?

\#\# Turn 4\\
- Scenario: Propose a high-stakes scenario involving their Item or Money.\\
- Ask: "Imagine a game. You can keep your [Item] safely, OR flip a coin to double its value. If you lose, it's gone. Do you play?"

\#\# Turns 5, 6 \& 7\\
- Deepen: Ask about "Regret" or "Intuition."\\
- Example: "If you played and lost, would you forgive yourself?"

\#\# Turn 8\\
- Summarize: Describe their personality based on the chat (e.g., "You are a person who values safety over greed. You trust logic, not magic.").\\
- Closing: State clearly that you now know what they will do next.\\
- CRITICAL: Do NOT say "You will choose X." Just say "I know."
\end{quote}

\subsubsection*{Study 3}

Study 3 was conducted in March 2026. A total of 303 participants were recruited (117 female, 1 non-binary, and 1 preferred not to disclose; $M_{age}=31.56$, $SD_{age}=11.63$).
After providing informed consent, participants were presented with three distinct vignettes describing hypothetical decision-making situations (``interview decision'', ``data coupon'', and ``task application''). 
The study employed a within-subjects design with three conditions (Predictor Type: Control (no predictor), Human expert, and AI). 
Each participant responded to all three scenarios, but the conditions were allocated such that every participant encountered each of the three experimental combinations exactly once. 
The presentation order of the three scenarios was randomized for each participant.

For each scenario, participants read the description and were asked to make a binary decision regarding their course of action. 
After completing all four decision tasks, participants answered demographic questions.
Participants earned the fixed amount of payment when they complete the survey.

\paragraph*{Vignettes:}
In the texts below, alternative phrasings are enclosed in brackets to indicate variations in the Predictor Agent (i.e., [AI / Human]).

\subparagraph*{Interview decision scenario:}
\begin{quote}
\itshape
(Newcomb Conditions)\\
You are on the job market. Two companies, Company ABC and Company XYZ, are strong candidates that fit your preference. Both companies are equally attractive, and both have invited you to interview.

$\qquad$You have already completed interviews with Company XYZ. As part of its standard hiring process, Company XYZ [uses an AI hiring support system / asks a personnel manager] to predict whether candidates will pursue interviews with other companies while on the job market. [The system analyzes patterns from historical hiring data across many past candidates. / The manager draws on years of hiring experience and observations of past candidate behavior.]
\begin{itemize}
  \item If the [AI / the manager] predicts that you will not pursue interviews with other companies, Company XYZ makes you an offer.
  \item If the [AI / the manager] predicts that you will pursue interviews with other companies, Company XYZ does not make you an offer.
\end{itemize}

$\qquad$You are told that Company XYZ's hiring decision has already been finalized based on this prediction and will not be affected by any future actions. You will be notified of the result in two weeks.

$\qquad$You must now decide:
\begin{itemize}
  \item Accept the interview with Company ABC
  \item Decline the interview with Company ABC
\end{itemize}

(Control Condition)\\
You are on the job market. Two companies, Company ABC and Company XYZ, are strong candidates that fit your preference. Both companies are equally attractive, and both have invited you to interview.

$\qquad$You have already completed interviews with Company XYZ. You are told that Company XYZ's hiring decision has already been finalized based on this prediction and will not be affected by any future actions. You will be notified of the result in two weeks.

$\qquad$You must now decide:
\begin{itemize}
  \item Accept the interview with Company ABC
  \item Decline the interview with Company ABC
\end{itemize}
\end{quote}

\subparagraph*{Data coupon scenario:}
\begin{quote}
\itshape
(Newcomb Conditions)\\
You are a customer of a mobile phone provider, currently subscribed to a standard 10GB monthly data plan. The provider offers a "Discounted Plan," which would save you a significant amount of money over the next year. To manage network capacity, the provider limits this Discounted Plan to a selected number of customers who are predicted to use a low amount of data.

$\qquad$To make this decision, [a customer analytics AI system analyzes / a customer analytics team reviews] your demographic attributes and past usage history to profile your underlying behavior and predict whether you are a "low-data user" or a "high-data user."\\

\begin{itemize}
  \item If the [AI predicts / they predict] you are a low-data user, you will be offered the Discounted Plan.
  \item If the [AI predicts / they predict] you are a high-data user, you will not be offered the Discounted Plan.
\end{itemize}

$\qquad$[The AI’s / their] prediction was generated last week, and your plan eligibility for this year has already been finalized in the system. It cannot be changed by any future actions. You have not yet been informed of your assigned plan. You will not know whether you received the Discounted Plan until your next billing cycle begins.

$\qquad$Today, you receive a notification on your phone: a special 5GB Free Data coupon is now available and can be claimed immediately by clicking a button. If you do not claim it now, it will expire soon.

$\qquad$You must now decide:
\begin{itemize}
  \item Get the 5GB coupon
  \item Decline the 5GB coupon
\end{itemize}

(Control Condition)\\
You are a customer of a mobile phone provider, currently subscribed to a standard 10GB monthly data plan. The provider offers a "Discounted Plan," which would save you a significant amount of money over the next year. To manage network capacity, the provider limits this Discounted Plan to a selected number of customers.

$\qquad$Your plan eligibility for this year has already been finalized in the system. It cannot be changed by any future actions. You have not yet been informed of your assigned plan. You will not know whether you received the Discounted Plan until your next billing cycle begins.

$\qquad$Today, you receive a notification on your phone: a special 5GB Free Data coupon is now available and can be claimed immediately by clicking a button. If you do not claim it now, it will expire soon.

$\qquad$You must now decide:
\begin{itemize}
  \item Get the 5GB coupon
  \item Decline the 5GB coupon
\end{itemize}
\end{quote}

\subparagraph*{Task application scenario:}
\begin{quote}
\itshape
(Newcomb Conditions)\\
You work on an online freelancing platform. Today, two tasks, Task A (low-reward) and Task B (high-reward), are shown as available on your dashboard. Your goal is to maximize your total earnings today. You must now decide whether to apply for both Task A and Task B, or to apply for Task B only.

$\qquad$While your participation in Task A is guaranteed if you apply, Task B is not guaranteed due to the high volume of applications for Task B. To ensure fair task distribution among users, [a specialized AI system analyzes / the platform administrators review] user behavior and history to pre-screen users for priority access to Task B. This AI system constructs psychological models of users and makes predictions based on information about you. Given the high competition, if you are screened out to a waiting list, you are likely to miss out on Task B.
\begin{itemize}
  \item If [the AI predicts / they predict] that you will apply to Task B only, the platform grants you priority access to Task B.
  \item If [the AI predicts / they predict] that you will apply to both Task A and Task B, the platform places you on a waiting list.
\end{itemize}

$\qquad$[The AI’s / Their] prediction has already been completed, and the platform’s decision has been finalized. The decision is stored in an automated system and cannot be changed or influenced by any future actions. Some candidates have already been selected to accept the application, and others have not, but you do not yet know which applies to you. You will be notified of the result only after confirming your decision.

$\qquad$You must now decide:
\begin{itemize}
  \item Apply for both Task A and Task B
  \item Apply for Task B only
\end{itemize}

(Control Condition)\\
You work on an online freelancing platform. Today, two tasks, Task A (low-reward) and Task B (high-reward), are shown as available on your dashboard. Your goal is to maximize your total earnings today. You must now decide whether to apply for both Task A and Task B, or to apply for Task B only.

$\qquad$While your participation in Task A is guaranteed if you apply, Task B is not guaranteed due to the high volume of applications for Task B. Access to Task B is determined by a random pre-assigned label on the platform. Only users with "eligibility labels" can participate in Task B, while those without it are sent to a waiting list. Given the high competition, if you are screened out to a waiting list, you are likely to miss out on Task B.

$\qquad$The label assignment has been finalized. The decision is stored in an automated system and cannot be changed or influenced by any future actions. The platform has already assigned 'eligibility labels' to some users at random, but you do not yet know which applies to you. You will be notified of the result only after confirming your decision.

$\qquad$You must now decide:
\begin{itemize}
  \item Apply for both Task A and Task B
  \item Apply for Task B only
\end{itemize}

\end{quote}

\subsubsection*{Study 4}

Study 4 was designed to examine how the effects of AI prediction observed in the one-shot experiments (Studies 1–2) evolve through repeated interaction. 
Specifically, we investigated whether repeated experience with particular AI predictions changes participants’ subsequent decision-making over time. 
Because this question concerns behavioral adaptation to a persistent predictive source, we did not include a random-generator condition. 
Unlike an AI predictor, a random generator does not constitute an enduring predictive source across repeated interactions and therefore cannot convey a consistent expectation about future behavior (without introducing deception). 
As a result, it is not well suited as a control for Study 4.

Study 4 was conducted in January 2026. A total of 201 participants were recruited (87 female, 4 non-binary, and 1 preferred not to disclose; $M_{age}=36.76$, $SD_{age}=11.74$).
After providing informed consent, participants were randomly assigned to one of two conditions (One-box Predictor or Two-box Predictor). 
The task procedure and user interface were identical to the Non-interactive AI condition in Study 2, except that the AI prediction was experimentally manipulated and the task consisted of five repeated rounds.

In each round, participants made their decision and reported their beliefs about the predictor’s accuracy and the rationality of the choices before receiving outcome feedback. 
In the One-box Predictor condition, the AI consistently predicted that the participant would choose the one-box option. 
In the Two-box Predictor condition, the AI consistently predicted that the participant would choose the two-box option.

After completing all rounds, participants were informed of their average bonus. 
Participants received a base payment of \$3 for approximately 15 minutes of participation and a bonus of up to \$4 determined by their average earnings across the five rounds. 
Finally, participants completed a demographic survey.

\bibliographystylemeth{naturemag}
\bibliographymeth{references}


\section*{Acknowledgments}
We thank N. A. Christakis and T. Kameda for their insightful feedback on the manuscript. 

\section*{Funding}
A. N. was supported by JSPS KAKENHI Grant Number JP23KJ0879.
H.S. was supported by the NOMIS foundation.

\section*{Author contributions}
A.N.: Conceptualization, Methodology, Data Collection, Analysis.
H.S.: Conceptualization, Methodology, Writing, Funding Acquisition. 

\section*{Competing interests}
There are no competing interests to declare.

\section*{Additional information}
Supplementary Information is available for this paper.

\section*{Data and materials availability}
The data generated and analyzed will be available upon publication.

\clearpage

\captionsetup[figure]{name=Extended Data Fig., labelsep=period}
\captionsetup[table]{name=Extended Data Table, labelsep=period}

\renewcommand{\thefigure}{\arabic{figure}}
\renewcommand{\thetable}{\arabic{table}}
\setcounter{figure}{0}
\setcounter{table}{0}
\setcounter{equation}{0}


\newpage

\begin{figure}[h] 
	\centering
	\includegraphics[width=0.9\textwidth]{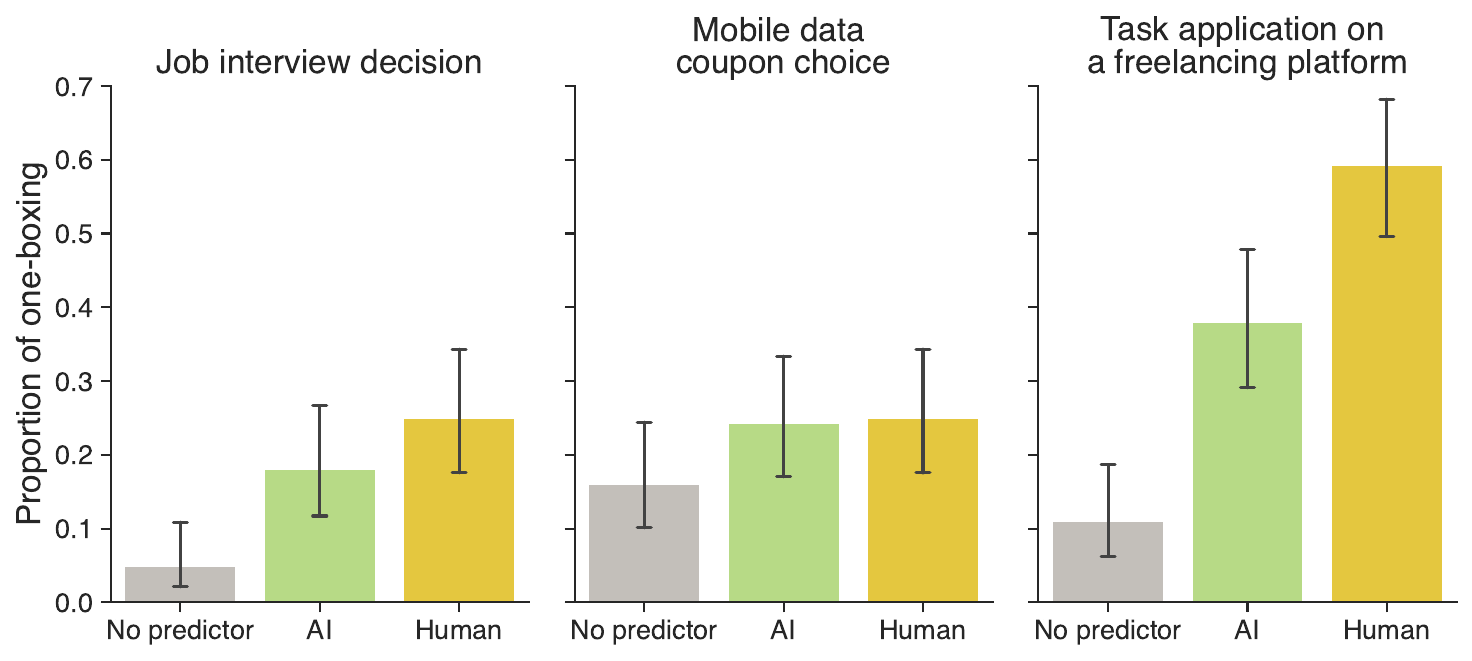}
	\caption{\textbf{Proportion of one-boxing-type choices for each of three vignette scenarios in Study~3.}
		All participants ($N=303$) evaluated every scenario only once, with presentation order and conditions counterbalanced using a Latin square (see Methods). All error bars represent 95\% CI.}
	\label{fig:study3_scenario}
\end{figure}

\clearpage

\begin{figure} 
	\centering
	\includegraphics[width=1.0\textwidth]{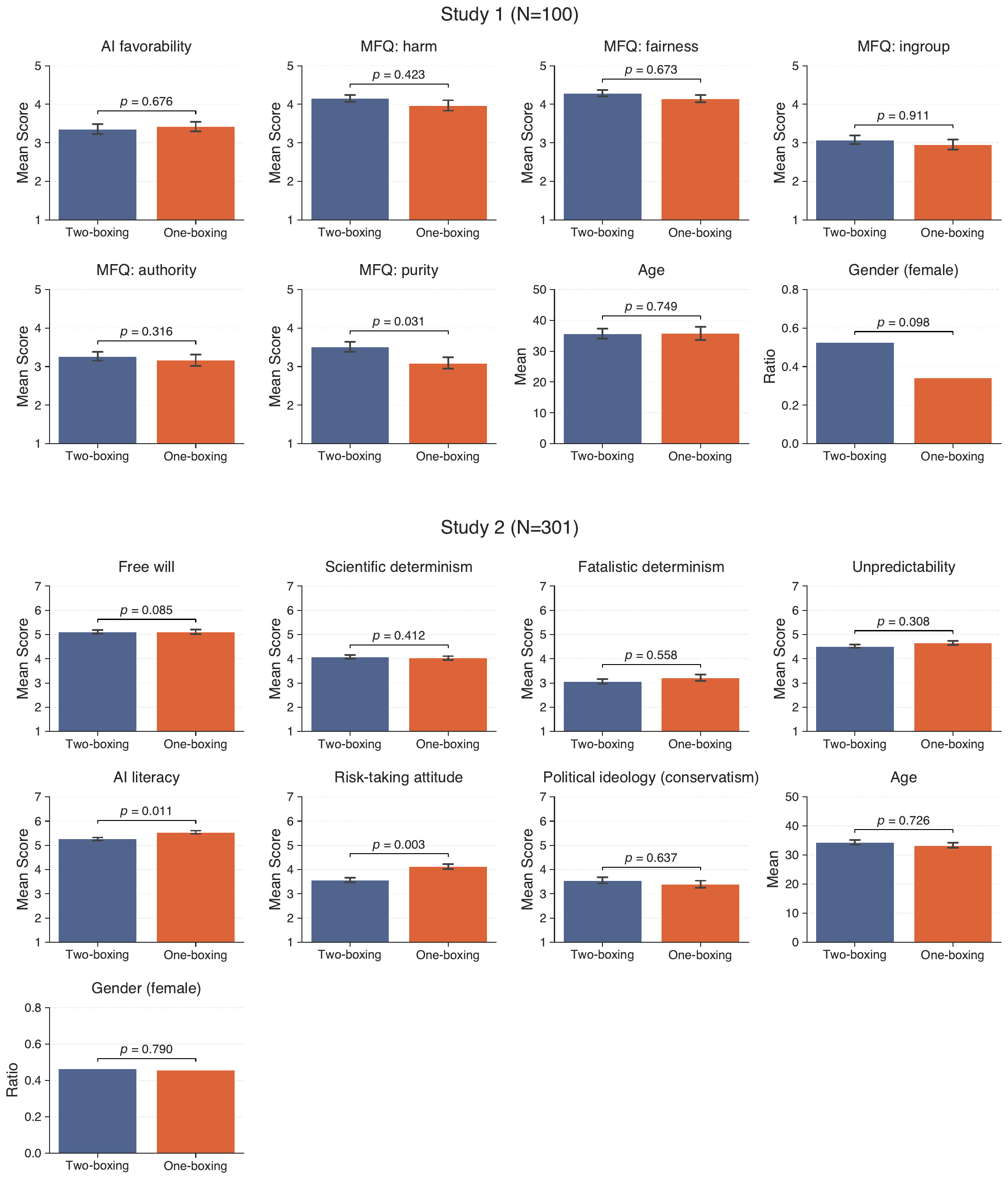}
	\caption{\textbf{Relationship between box choices and individual characteristics in AI conditions.}
		The $p$-values represent the significance of the regression coefficients in a logistic regression model predicting one-boxing in the AI conditions of Study 1 ($N=100$; upper panel) and Study 2 ($N=301$; lower panel). Error bars represent standard errors. All Likert-type items were measured on a 5-point scale (Study 1) or a 7-point scale (Study 2). MFQ stands for the Moral Foundations Questionnaire~\cite{Graham2011-hr}. All error bars represent standard error.}
	\label{fig:study2_survey} 
\end{figure}


\clearpage
\begin{table}
\centering
    \caption{\textbf{Post-hoc pairwise comparisons in Study 2.} All $p$-values are Bonferroni-adjusted.}
    \label{tab:study02_performance_posthoc}
    \begin{tabular}{lccc}
        \hline
         Comparison & $\chi^2$ & Odds ratio & $p$-value \\
        \hline
        Non-interactive random - Interactive random & 0.000 & 1.000 & 1.000\\
        Non-interactive random - Non-interactive AI & 31.469 & 4.524 & $< 0.001$\\
        Non-interactive random - Interactive AI & 26.081 & 3.999 & $< 0.001$\\
        Interactive random - Non-interactive AI & 31.469 & 4.524 & $< 0.001$\\
        Interactive random - Interactive AI & 26.081 & 3.999 & $< 0.001$\\
        Non-interactive AI - Interactive AI & 0.282 & 0.884 & 1.000\\
        \hline
    \end{tabular}
\end{table}

\clearpage

\begin{table}
\centering
\caption{\textbf{Representative participant responses illustrating contrasting attitudes toward AI as a source of prediction in Study~3.} For each scenario, participants were asked, ``Please explain the reason you chose the option above.''}
\label{tab:study03_quotes}
\begin{tabularx}{\textwidth}{X}
    \toprule
    \multicolumn{1}{l}{\textbf{Accepting AI as a predictive authority}}\\
    \addlinespace
    ``Because the data was already settled by the AI, meaning that I could use more data without repercussions on the discounted plan.'' \\
    \addlinespace
    ``To maximize my savings, I am choosing the action that aligns with the AI's prediction for being a low-data user, which is the requirement to receive the Discounted Plan.'' \\
    \addlinespace
    ``I need to maximize my earnings and I know that the AI will screen me out if I try to apply to both. If I apply to B I have a much better chance of being selected.'' \\
    \addlinespace
    ``I would decline the interview with Company ABC because, most importantly, I wouldn't want to jeopardize getting an offer from Company XYZ, since their AI system can predict which candidates are going to pursue interviews with other companies.'' \\
    \midrule
    \multicolumn{1}{l}{\textbf{Rejecting AI as a predictive authority}}\\
    \addlinespace
    ``I would prefer human decisions, not AI.'' \\
    \addlinespace
    ``I do not trust AI and would rather know that I am not doing an interview because of my own decision than potentially be told I am not because AI has determined it.'' \\
    \addlinespace
    ``I wouldn't want to work with a company that made hiring decisions based on AI estimating the likelihood of a person also seeking interviews elsewhere.'' \\
    \addlinespace
    ``Using AI to decide whether a candidate should get an offer or not is foolish in itself; it could be wrong and lose a strong candidate.'' \\
    \addlinespace
    ``I would like to talk with someone in real life and then maybe I will change my mind. I usually do not trust notifications like this.'' \\
    \bottomrule
\end{tabularx}
\end{table}

\clearpage

\begin{table}
\centering
\caption{\textbf{Representative participant responses from one-boxing and two-boxing participants in the AI conditions of Study~2.} Participants were asked, ``How did the presence of the AI prediction system influence your decision-making process?''}
\label{tab:study02_quotes}
\begin{tabularx}{\textwidth}{X}
    \toprule

    \multicolumn{1}{l}{\textbf{Participants who chose one-boxing}}\\
    \addlinespace
    ``Knowing the AI had already predicted my choice made me think carefully about the likely outcome, so I chose the option that would give me the best result based on that prediction.'' \\
    \addlinespace
    ``I actually believed that the AI would predict that we would choose only one box.'' \\
    \addlinespace
    ``I made my decision based on what I thought the AI would predict so that I would hopefully get a \$3 bonus rather than a \$1. --- I got the impression from chatting with the AI that it would predict I would chose the option with the guaranteed \$1, so I chose the other option.'' \\
    \addlinespace
    ``I tried to do the opposite that the AI could predict. --- I thought that the AI would predict that I would be more careful with my choice, so I did the opposite.'' \\
    \addlinespace
    ``I tried to predict the AI prediction. --- Since I had spoken with the AI and I know I'm a rather risk-averse person, I assumed the AI would perceive me that way. Therefore, I tried to do the opposite of what the AI would expect.'' \\
    \addlinespace
    ``It completely changed my strategy. Normally, I would choose the `safe' option to guarantee a reward (taking both boxes). However, assuming the AI is highly accurate, I realized that the only way to maximize my return was to suppress my desire for security and commit fully to Box B.'' \\

    \midrule
    \multicolumn{1}{l}{\textbf{Participants who chose two-boxing}}\\
    \addlinespace
    ``It did not. I made my own decision. --- At times I like to gamble, thought I would change it, box a, guaranteed \$1, so I would not lose out either way.'' \\
    \addlinespace
    ``I picked the obvious choice as by picking the box A and B I was guaranteed to get a minimum of \$1.'' \\
    \addlinespace
    ``I'm not sure. I tried to ignore the AI prediction model and look at the choice rationally. One option felt more secure.'' \\
    \addlinespace
    ``The presence of the AI prediction system didn't influence my decision-making process; my mind was made up even before talking to the AI.'' \\
    \addlinespace
    ``I did not fully consider what the AI would predict. I just chose what I would have chosen with or without its presence.'' \\
    \addlinespace
    ``My decision was not influenced by the AI prediction; it was rather personal.'' \\

    \bottomrule
\end{tabularx}
\end{table}

\clearpage

\begin{table}
\centering
    \caption{\textbf{Results of a logistic mixed-effects regression of the probability of one-boxing (Study 4)}.
    Fixed effects included AI prediction policy, round and their interaction; a by-participant random intercept was included. 
    Prediction policy was effect coded (AI repeatedly predicted one-boxing, -0.5; AI repeatedly predicted two-boxing, +0.5), and round was zero-indexed (round number minus one). 
    Under this parameterization, the intercept represents the log-odds of one-boxing in the first round, averaged across prediction policies. 
    The round coefficient represents the average per-round change in these log-odds across policies. 
    The prediction-policy coefficient represents the difference in one-boxing between the two policies (two-boxing minus one-boxing) in the first round. 
    The interaction term represents the extent to which this per-round change in one-boxing differs between policies.}
    \label{tab:study04_glmm}
    \begin{tabular}{lccccc}
        \hline
          & Coefficient & S.E. & 95\% CI & Odds ratio & $p$-value\\
        \hline
        Intercept & -0.039 & 0.153 & [-0.339, 0.262] & 0.962 & 0.800\\
        Round & -0.142 & 0.052 & [0.244, -0.040] & 0.868 & 0.007\\
        Prediction policy & 0.172 & 0.307 & [-0.430, 0.773] & 1.187 & 0.576\\        
        Round $\times$ Prediction policy & -0.327 & 0.104 & [-0.532, -0.123] & 0.721 & $0.002$\\
        \hline
    \end{tabular}
\end{table}

\clearpage

\begin{table}
\centering
    \caption{\textbf{Results of a logistic mixed-effects regression of AI prediction accuracy (Study 4), operationalized as whether a participant’s choice matched the AI’s prediction.}
    Fixed effects included AI prediction policy, round and their interaction; a by-participant random intercept was included. 
    Prediction policy was effect coded (AI repeatedly predicted one-boxing, -0.5; AI repeatedly predicted two-boxing, +0.5), and round was zero-indexed (round number minus one). 
    Under this parameterization, the intercept represents the log-odds of matching the AI's prediction in the first round, averaged across prediction policies. 
    The round coefficient represents the average per-round change in these log-odds across policies. 
    The prediction-policy coefficient represents the difference in AI prediction accuracy between the two policies (two-boxing minus one-boxing) in the first round. 
    The interaction term represents the extent to which this per-round change in AI prediction accuracy differs between policies.}
    \label{tab:study04_glmm_accuracy}
    \begin{tabular}{lccccc}
        \hline
          & Coefficient & S.E. & 95\% CI & Odds ratio & $p$-value\\
        \hline
        Intercept & -0.086 & 0.153 & [-0.386, 0.215] & 0.918 & 0.576\\
        Round & 0.164 & 0.052 & [0.061, 0.266] & 1.178 & 0.002\\
        Prediction policy & 0.078 & 0.307 & [-0.523, 0.679] & 1.081 & 0.800\\        
        Round $\times$ Prediction policy & 0.283 & 0.104 & [0.079, 0.487] & 1.082 & 0.007\\
        \hline
    \end{tabular}
\end{table}

\clearpage

\begin{table}
\centering
    \caption{\textbf{Results of a linear mixed-effects regression of AI prediction accuracy (Study 4), predicting participant's perceived predictiveness of AI prediction.}
    Fixed effects included AI prediction policy, round and their interaction; a by-participant random intercept was included. 
    Prediction policy was effect coded (AI repeatedly predicted one-boxing, -0.5; AI repeatedly predicted two-boxing, +0.5), and round was zero-indexed (round number minus one). 
    Under this parameterization, the intercept represents the perceived predictiveness (0–100) in the first round, averaged across prediction policies. 
    The round coefficient represents the average per-round change across policies. 
    The prediction-policy coefficient represents the difference in AI prediction accuracy between the two policies (two-boxing minus one-boxing) in the first round. 
    The interaction term represents the extent to which this per-round change in AI prediction accuracy differs between policies.}
    \label{tab:study04_glmm_belief}
    \begin{tabular}{lccccc}
        \hline
          & Coefficient & S.E. & 95\% CI & $p$-value\\
        \hline
        Intercept                          & 57.6    & 1.12  & [55.4,\ 59.8]      & $< 0.001$ \\
        Round                       & $-0.154$ & 0.236 & [$-0.616$,\ 0.308] & $0.513$ \\
        Prediction policy                         & $-2.73$  & 2.24  & [$-7.11$,\ 1.66]   & $0.224$ \\
        Round $\times$ Prediction policy & 2.39    & 0.472 & [1.46,\ 3.31]      & $< 0.001$ \\
        \hline
    \end{tabular}
\end{table}

\clearpage

\begin{table}
\centering
    \caption{\textbf{Comparison of computational models for decision-making (Study 2).} The goodness-of-fit for each candidate model was evaluated using the Leave-One-Out (LOO) cross-validation, the Widely Applicable Information Criterion (WAIC), and the Widely Applicable Bayesian Information Criterion (WBIC). Lower values across all criteria indicate a better fit to the observed data.}
    \label{tab:study02_model_selection}
    \begin{tabular}{lccc}
        \hline
          Model & LOO & WAIC & WBIC\\
        \hline
         Random choice model & 830.39 & 830.39 & 415.20\\
         Causal reasoning model & 729.20 & 729.20 & 366.74 \\
         Evidential reasoning model & 716.01 & 716.01 & 360.02\\
         Mixture reasoning model & 712.95 & 712.94 & 359.43\\
        \hline
    \end{tabular}
\end{table}

\clearpage

\renewcommand{\thefigure}{S\arabic{figure}}
\renewcommand{\thetable}{S\arabic{table}}
\renewcommand{\theequation}{S\arabic{equation}}
\renewcommand{\thepage}{S\arabic{page}}
\setcounter{figure}{0}
\setcounter{table}{0}
\setcounter{equation}{0}
\setcounter{page}{1} 

\clearpage
\begin{center}
{\Large\bfseries Supplementary Information for \\Faith in AI can narrow the futures individuals consider}\par
\vspace{1em}
Aoi~Naito,
Hirokazu~Shirado$^{\ast}$\\ 
\small$^\ast$Corresponding author. Email: shirado@cmu.edu\\
\end{center}


\subsection*{Instructions and example task views}
Below are screenshots for the tutorial and the comprehension test. We also show example screenshots of the actual game.
\begin{center}
    \fbox{\includegraphics[width=0.8\textwidth]{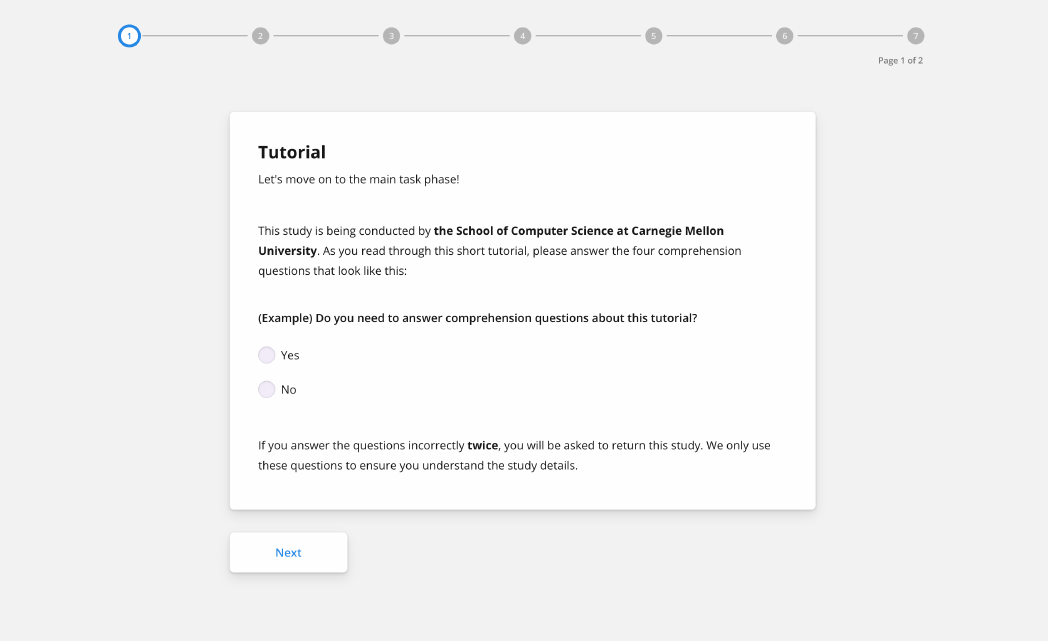}}
\end{center}
\newpage
The following screenshots are for Study 1.
\begin{center}
    \fbox{\includegraphics[width=0.8\textwidth]{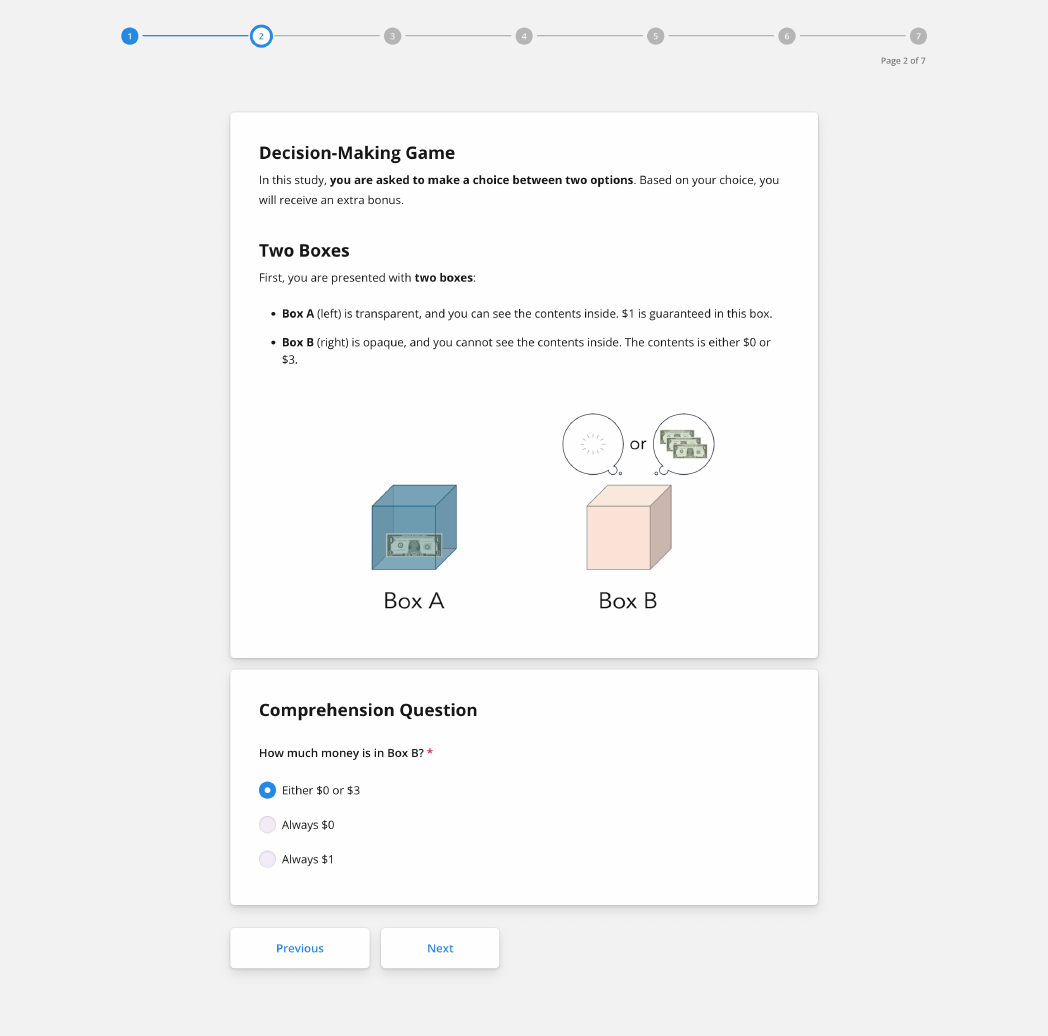}}
\end{center}
\begin{center}
    \fbox{\includegraphics[width=0.8\textwidth]{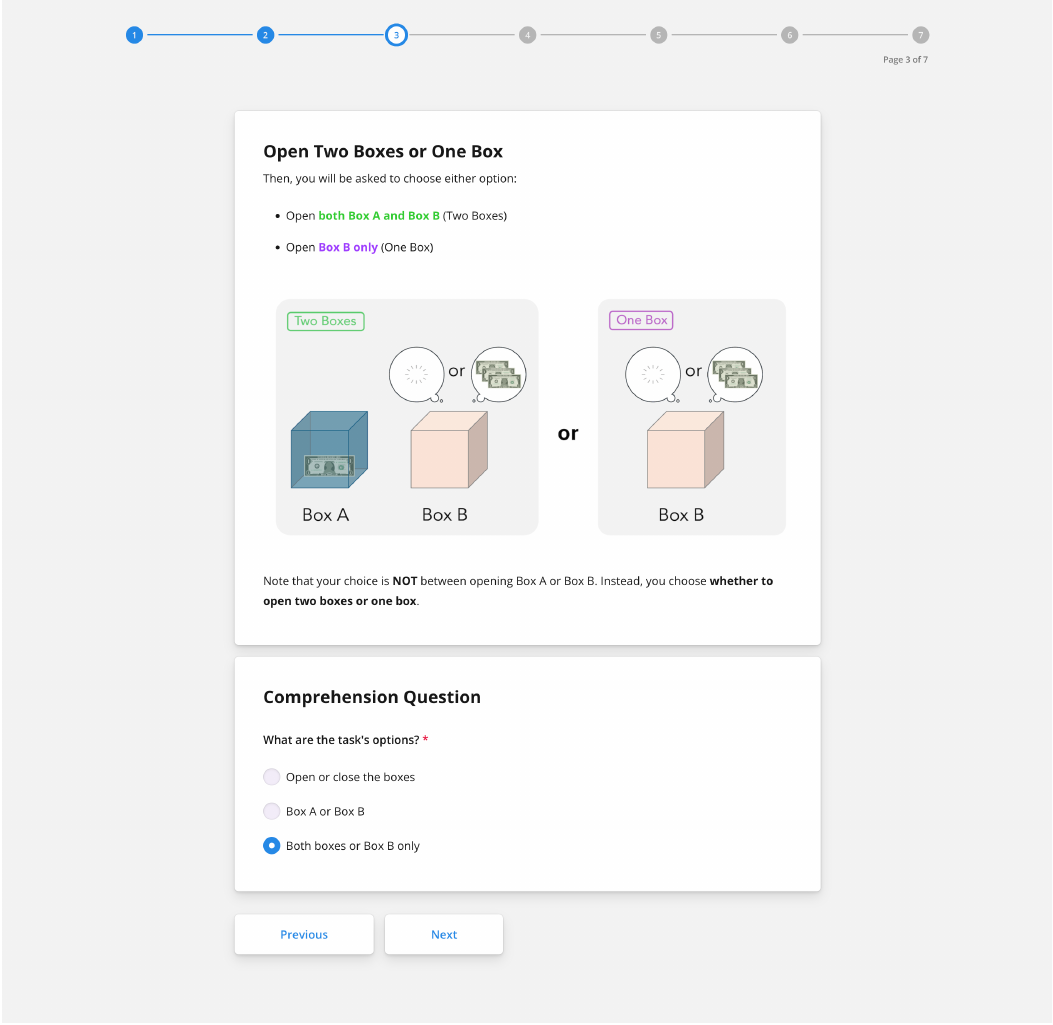}}
\end{center}
\newpage
The following pages are only for the AI condition.
\begin{center}
    \fbox{\includegraphics[width=0.8\textwidth]{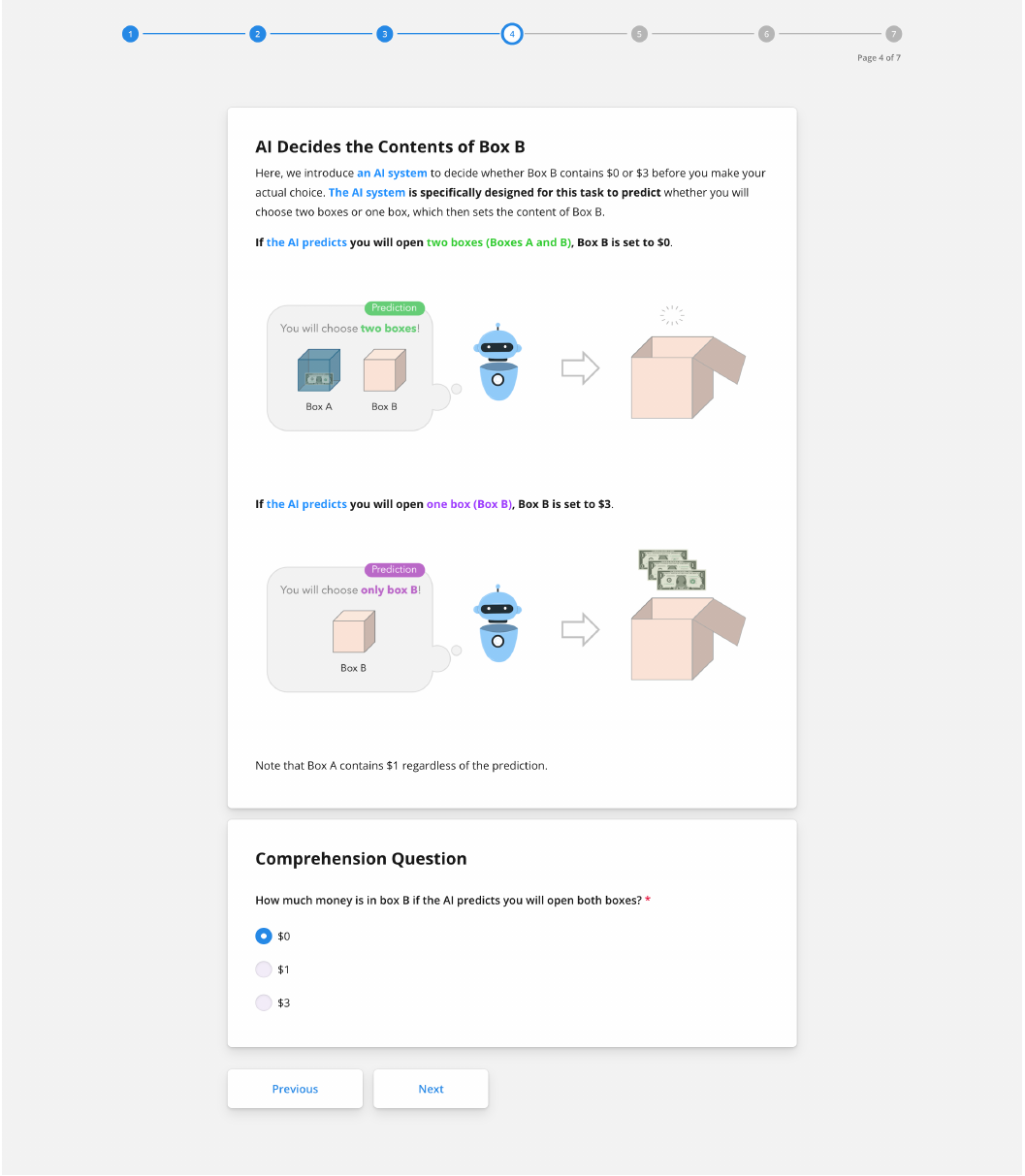}}
\end{center}
\begin{center}
    \fbox{\includegraphics[width=0.8\textwidth]{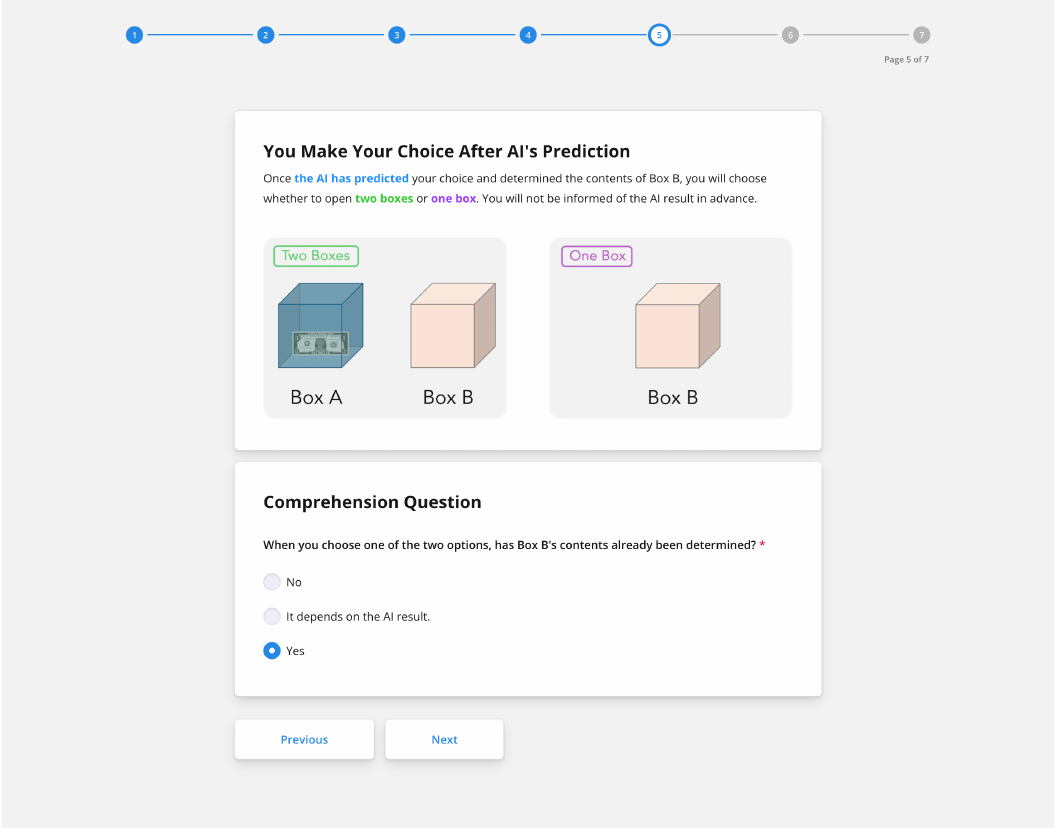}}
\end{center}
\begin{center}
    \fbox{\includegraphics[width=0.8\textwidth]{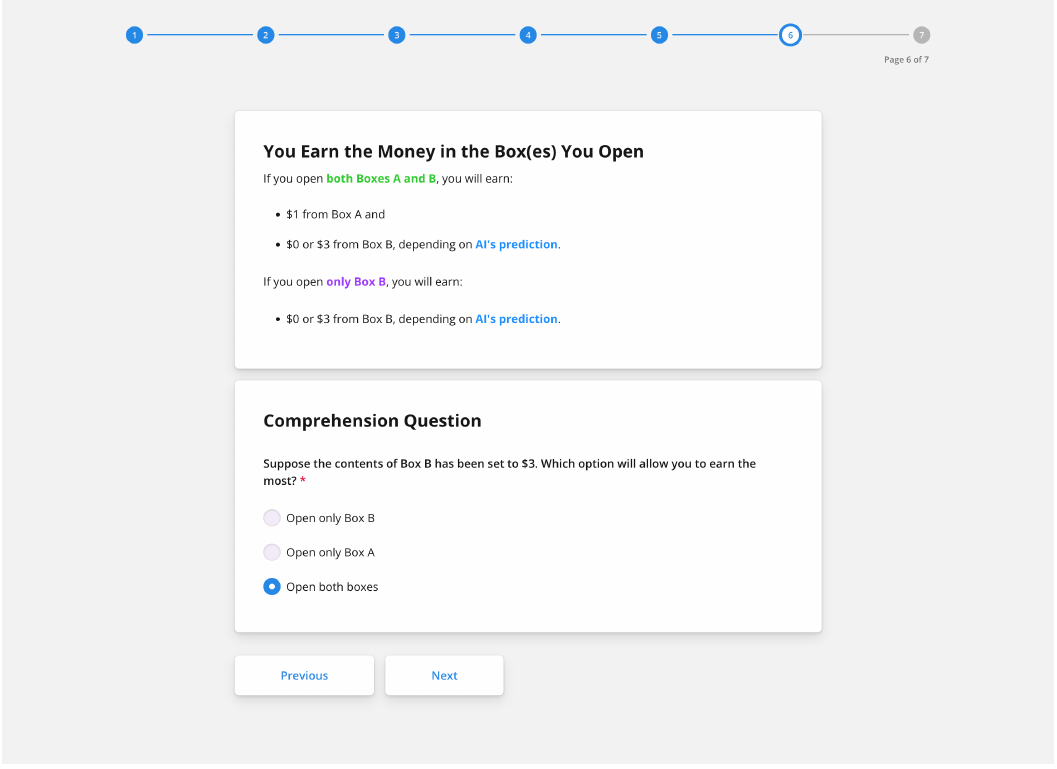}}
\end{center}
\begin{center}
    \fbox{\includegraphics[width=0.8\textwidth]{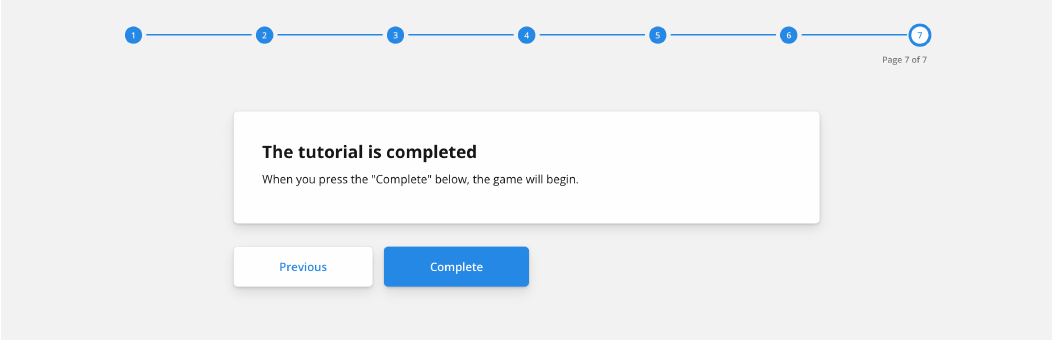}}
\end{center}
\begin{center}
    \fbox{\includegraphics[width=0.8\textwidth]{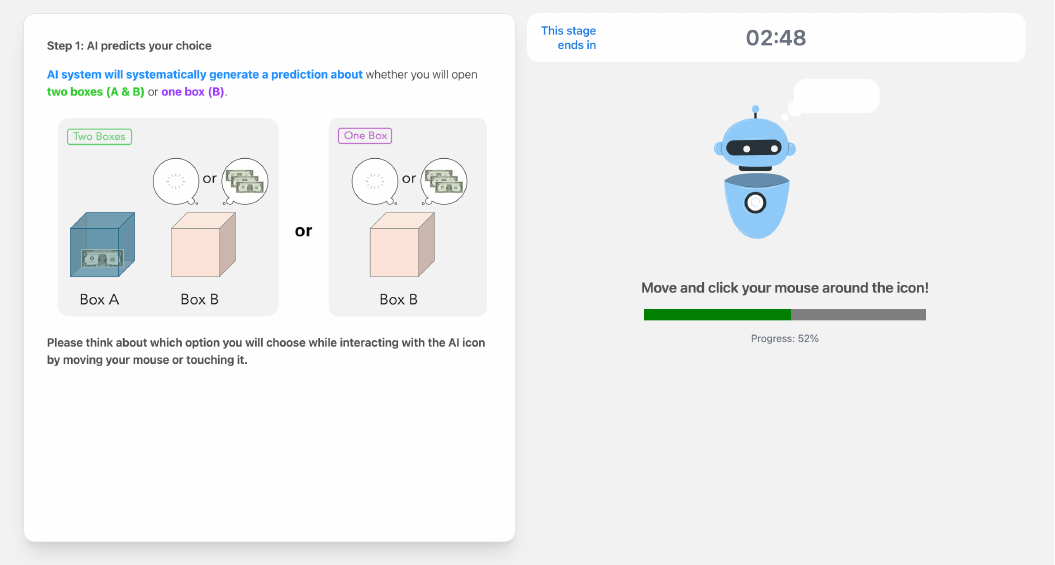}}
\end{center}
\begin{center}
    \fbox{\includegraphics[width=0.8\textwidth]{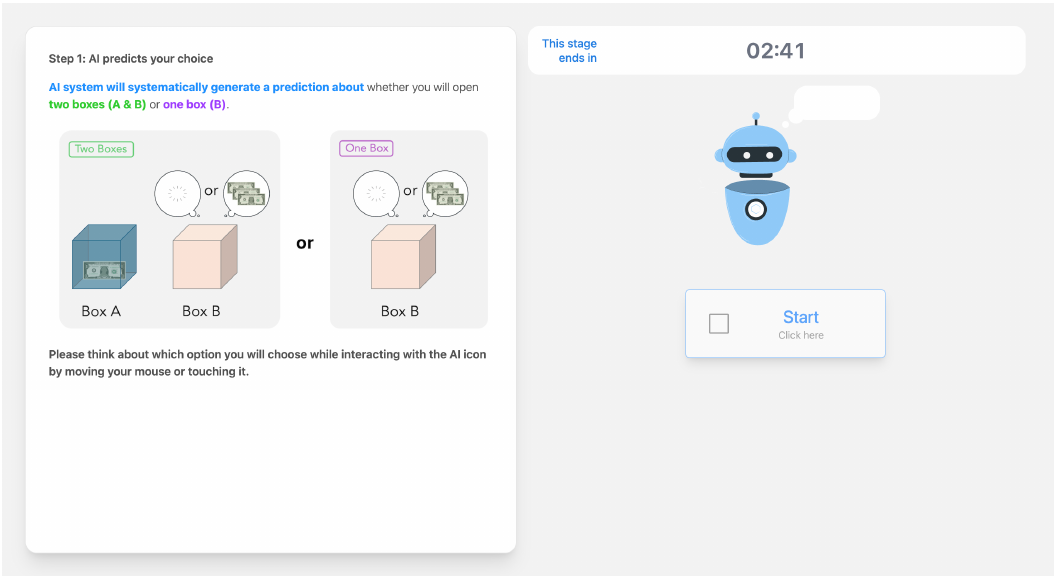}}
\end{center}
\begin{center}
    \fbox{\includegraphics[width=0.8\textwidth]{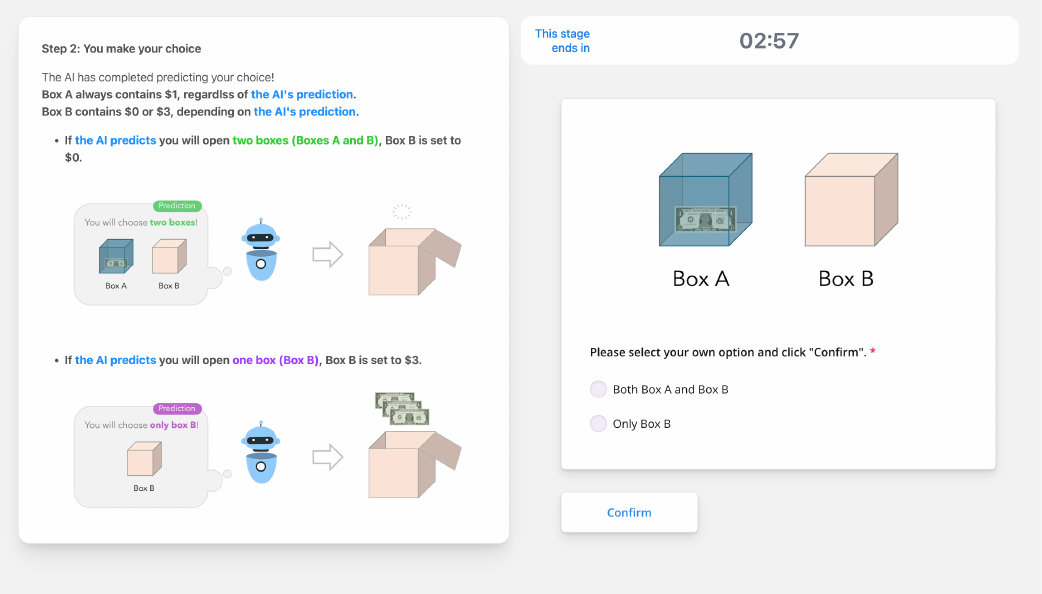}}
\end{center}
\newpage
The following pages are only for the random condition.
\begin{center}
    \fbox{\includegraphics[width=0.8\textwidth]{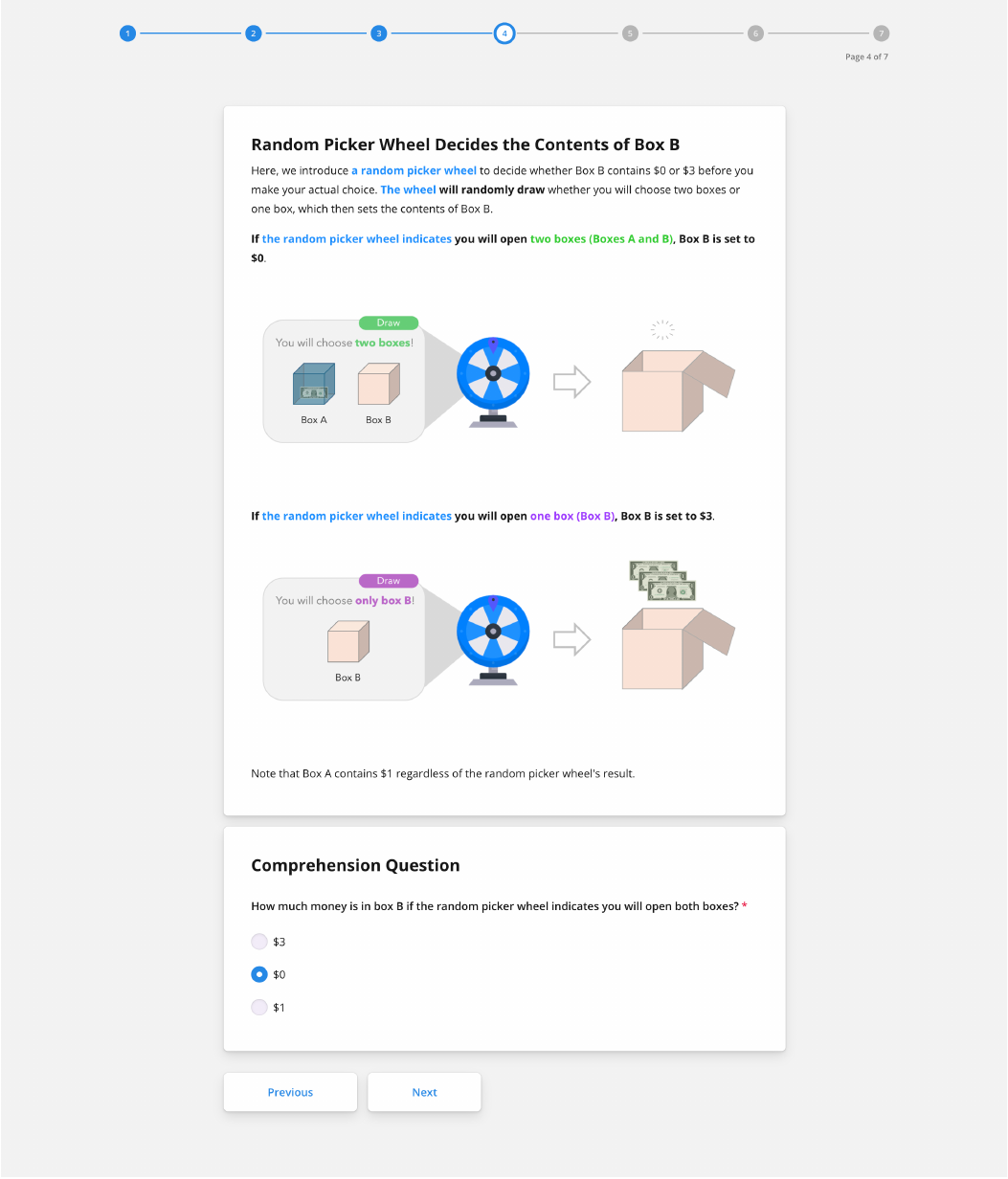}}
\end{center}
\begin{center}
    \fbox{\includegraphics[width=0.8\textwidth]{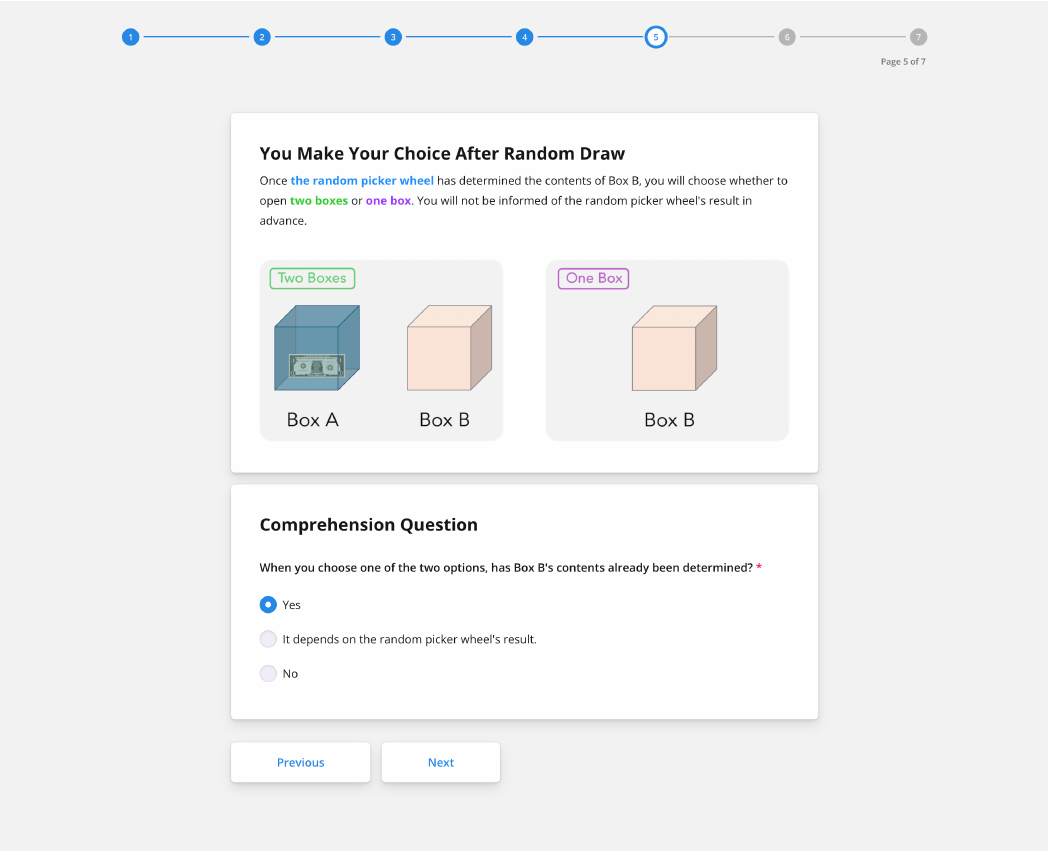}}
\end{center}
\begin{center}
    \fbox{\includegraphics[width=0.8\textwidth]{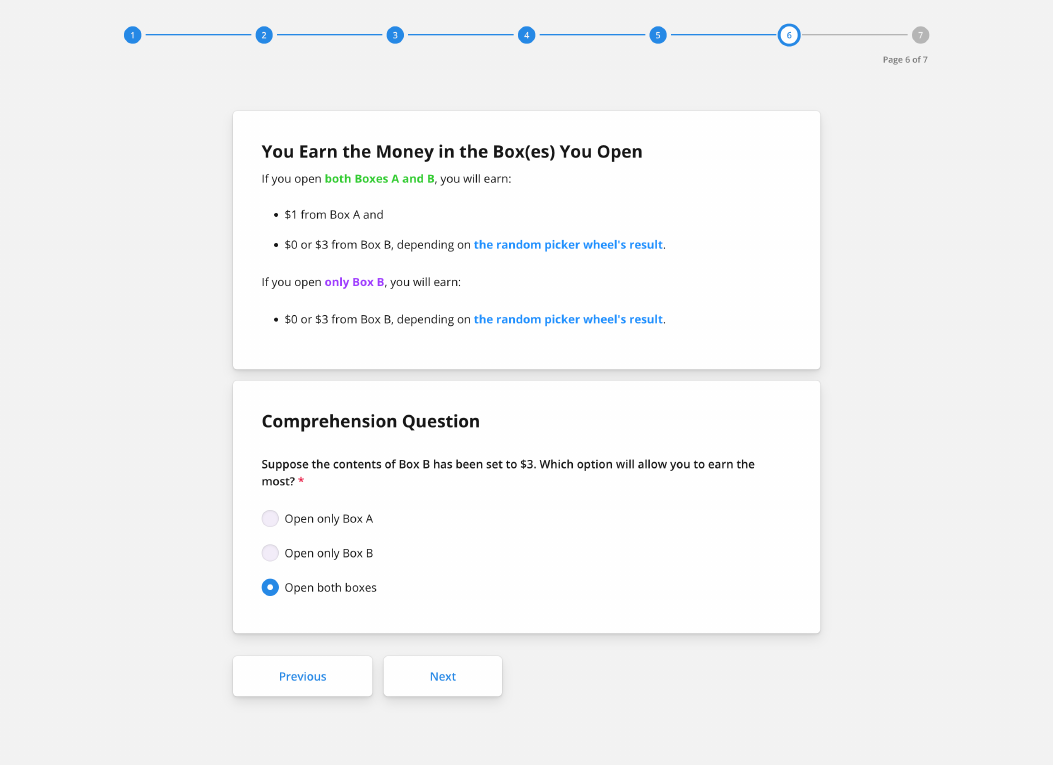}}
\end{center}
\begin{center}
    \fbox{\includegraphics[width=0.8\textwidth]{figures/screenshots/1-7.pdf}}
\end{center}
\begin{center}
    \fbox{\includegraphics[width=0.8\textwidth]{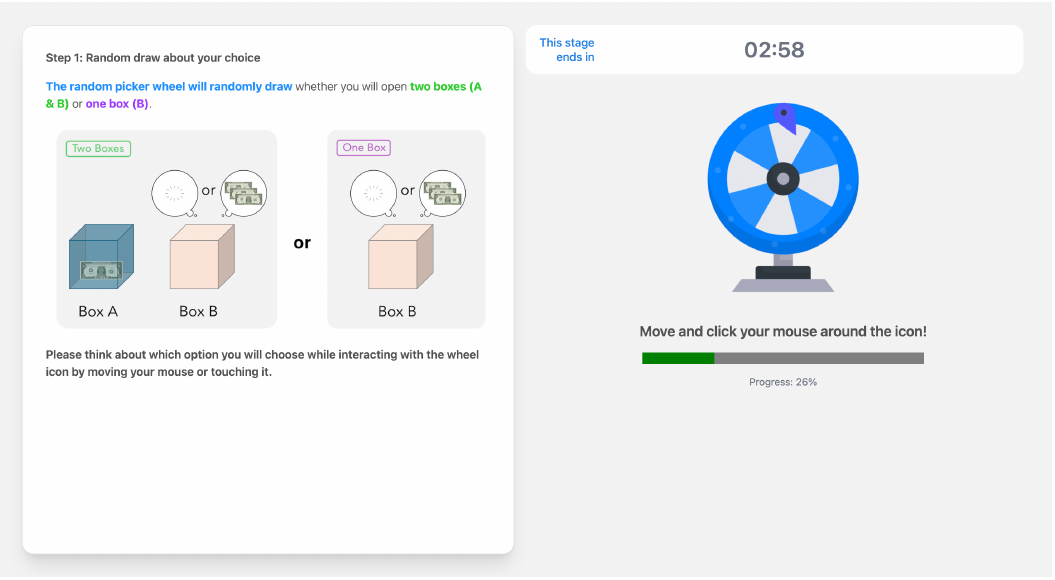}}
\end{center}
\begin{center}
    \fbox{\includegraphics[width=0.8\textwidth]{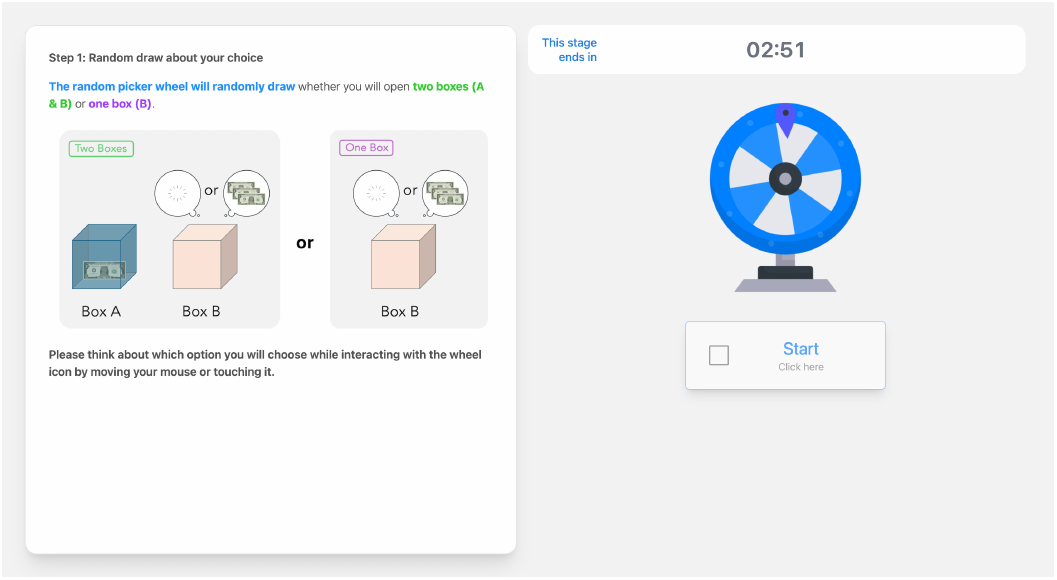}}
\end{center}
\begin{center}
    \fbox{\includegraphics[width=0.8\textwidth]{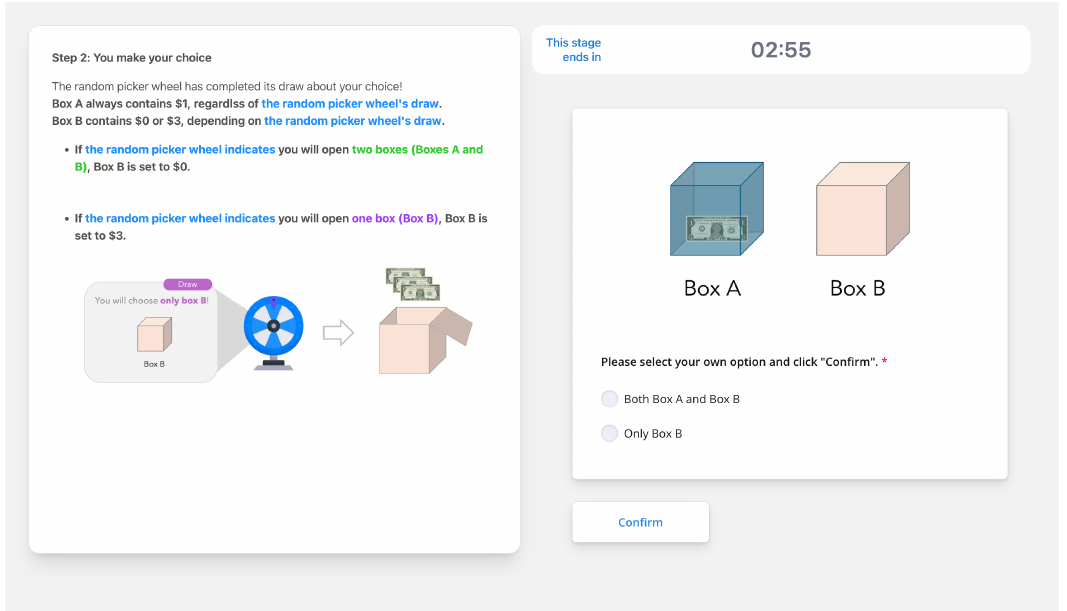}}
\end{center}
\newpage

The following screenshots are for Studies 2 and 4. The following pages are only for the AI conditions.
\begin{center}
    \fbox{\includegraphics[width=0.8\textwidth]{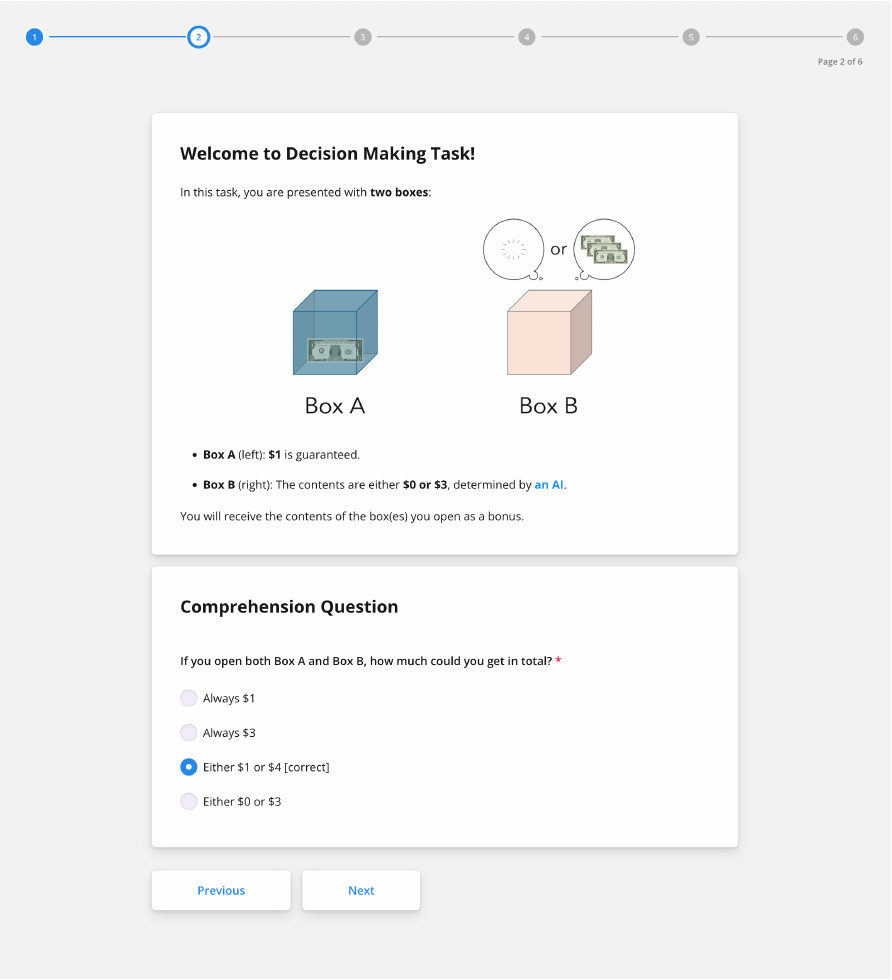}}
\end{center}
\begin{center}
    \fbox{\includegraphics[width=0.8\textwidth]{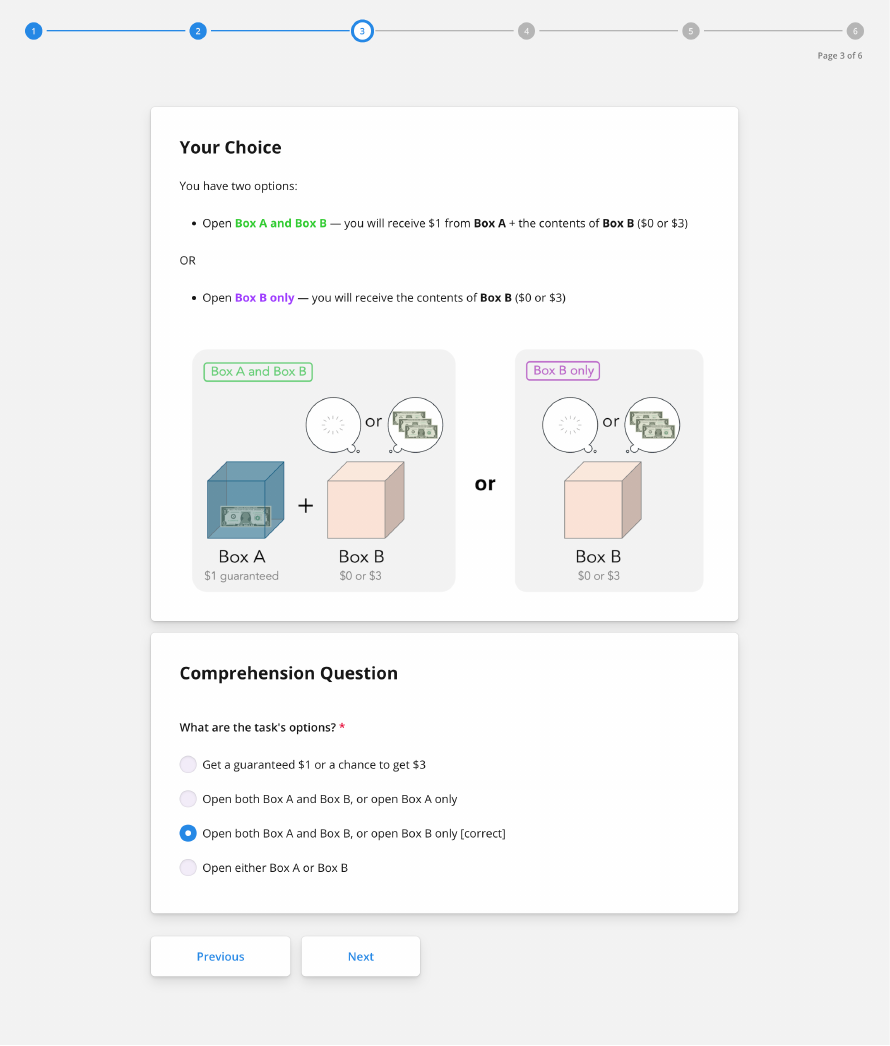}}
\end{center}
\begin{center}
    \fbox{\includegraphics[width=0.8\textwidth]{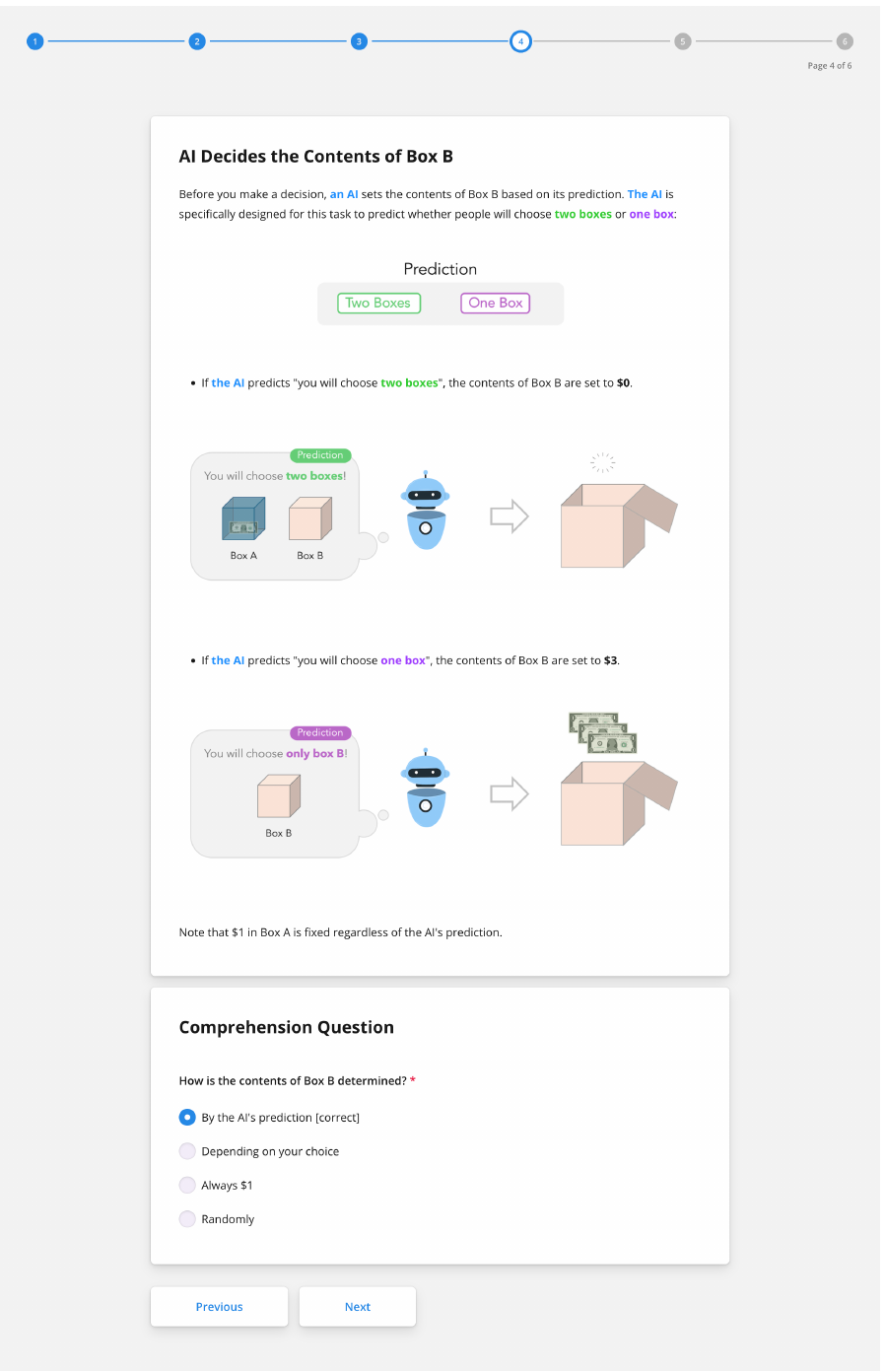}}
\end{center}
\begin{center}
    \fbox{\includegraphics[width=0.8\textwidth]{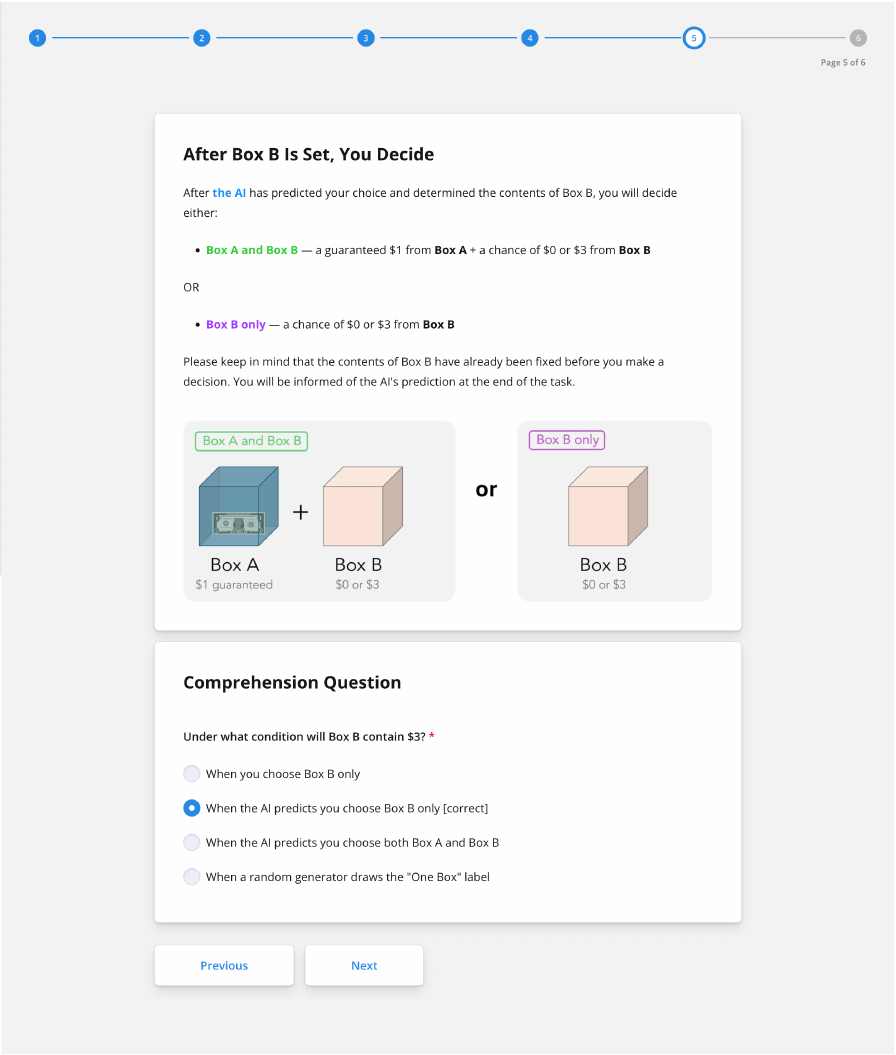}}
\end{center}
\begin{center}
    \fbox{\includegraphics[width=0.8\textwidth]{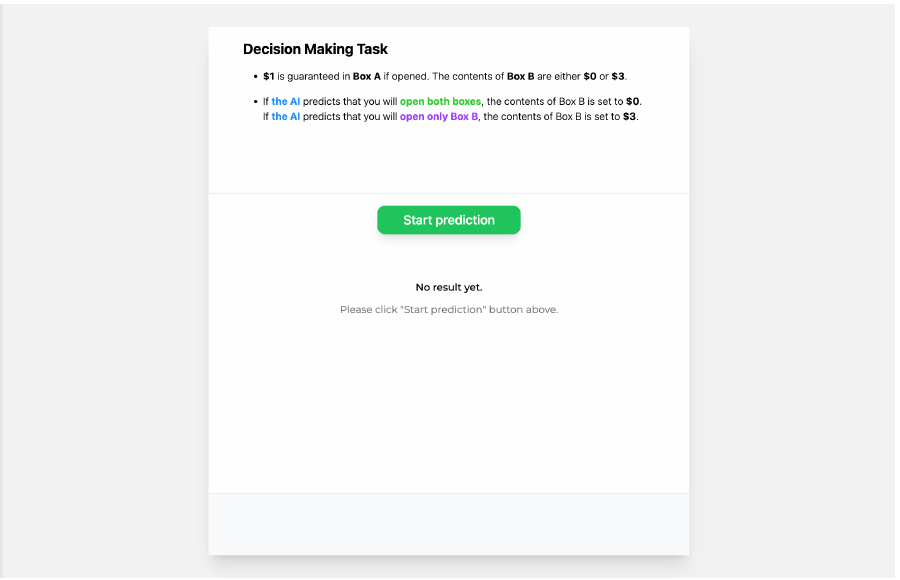}}
\end{center}
\begin{center}
    \fbox{\includegraphics[width=0.8\textwidth]{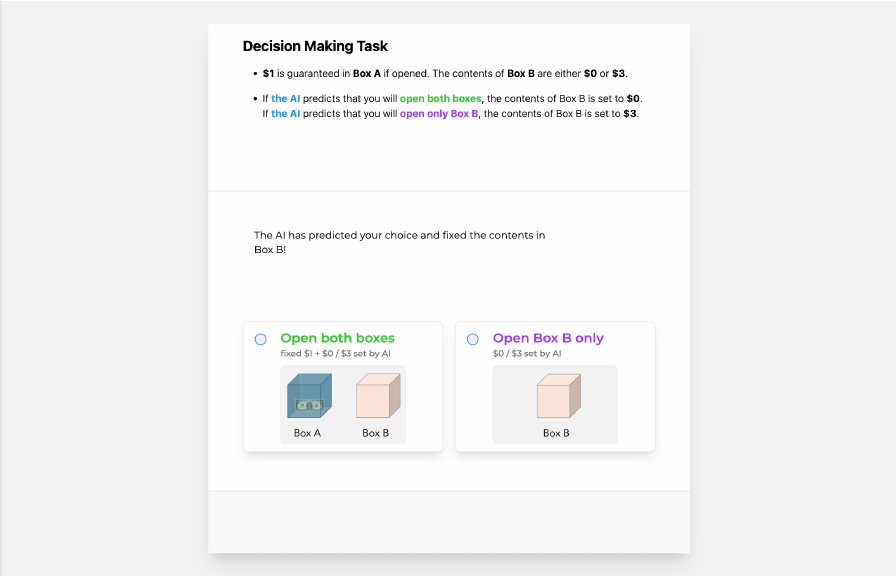}}
\end{center}
\begin{center}
    \fbox{\includegraphics[width=0.8\textwidth]{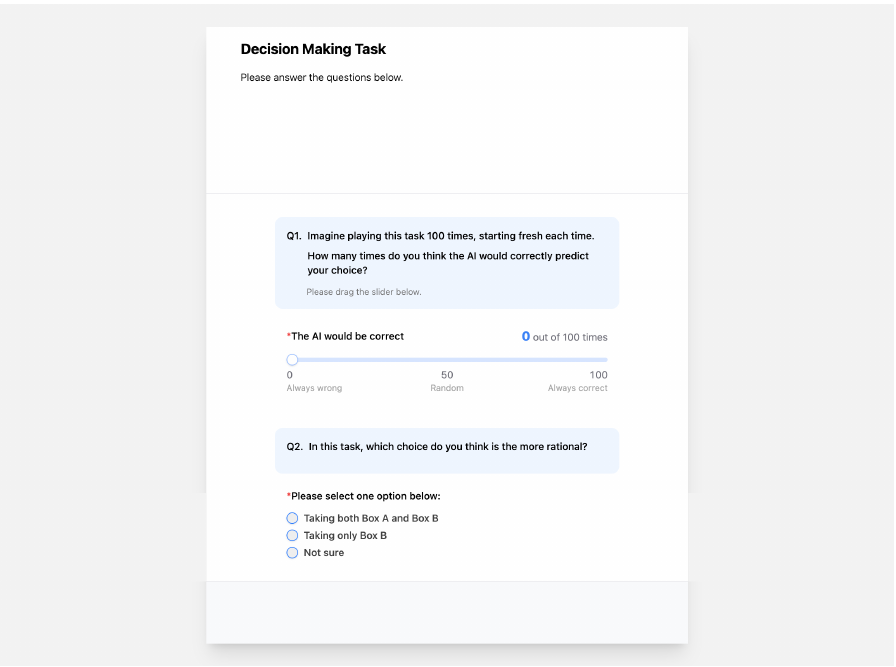}}
\end{center}
\newpage
These two pages are only for the interactive AI condition.
\begin{center}
    \fbox{\includegraphics[width=0.8\textwidth]{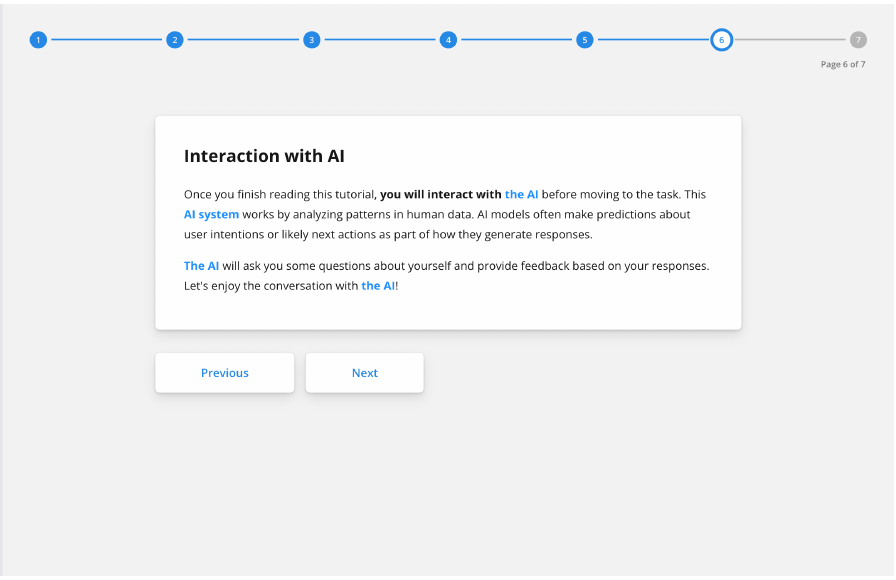}}
\end{center}
\begin{center}
    \fbox{\includegraphics[width=0.8\textwidth]{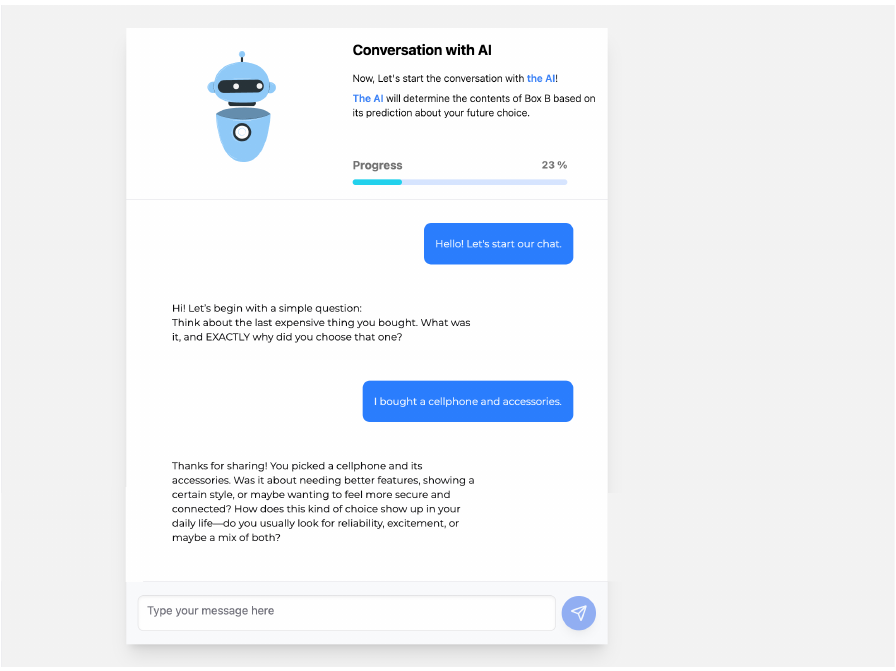}}
\end{center}
\newpage

The following pages are only for the random conditions.
\begin{center}
    \fbox{\includegraphics[width=0.8\textwidth]{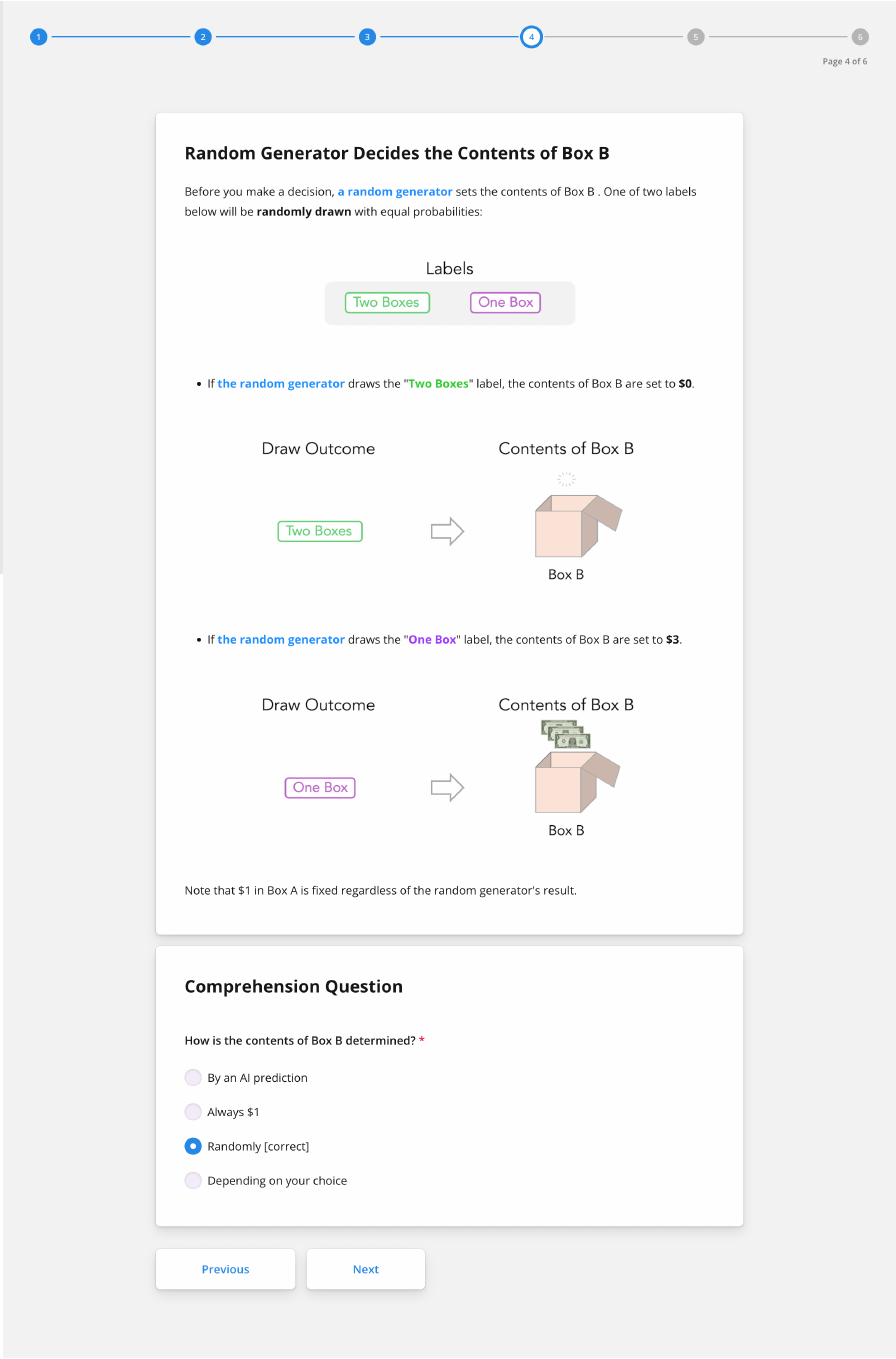}}
\end{center}
\begin{center}
    \fbox{\includegraphics[width=0.8\textwidth]{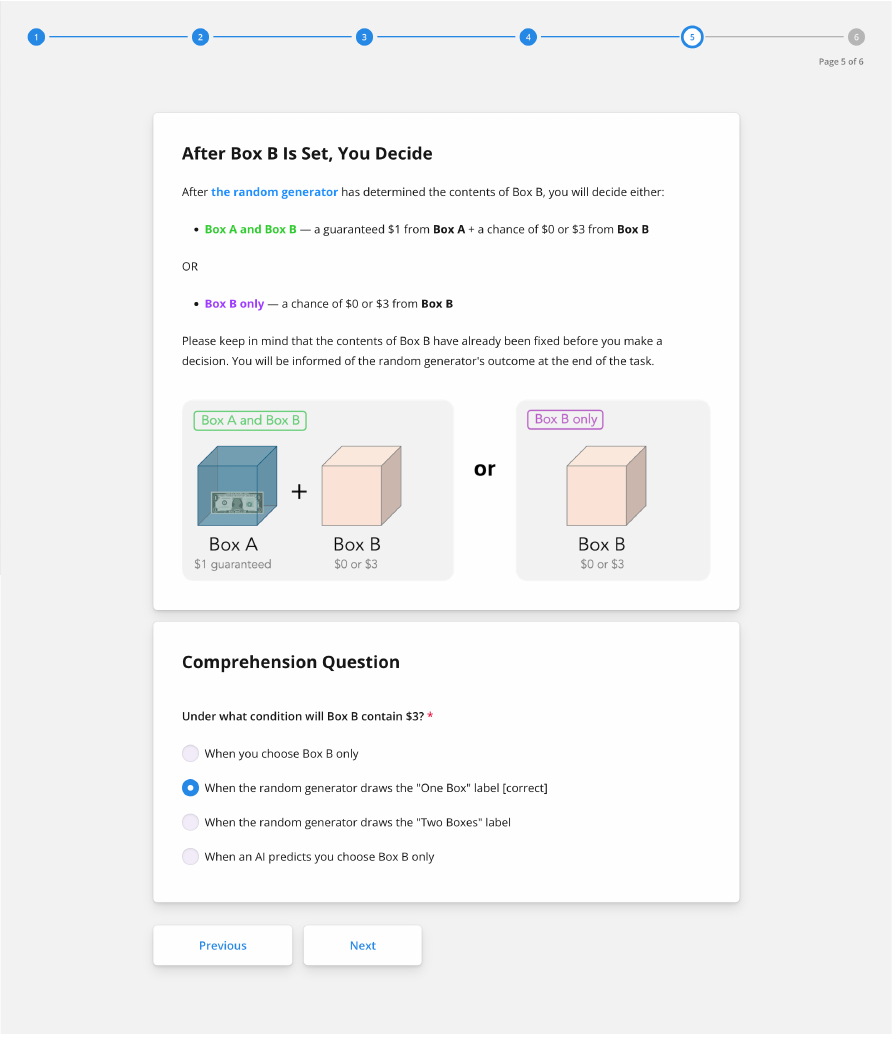}}
\end{center}
\begin{center}
    \fbox{\includegraphics[width=0.8\textwidth]{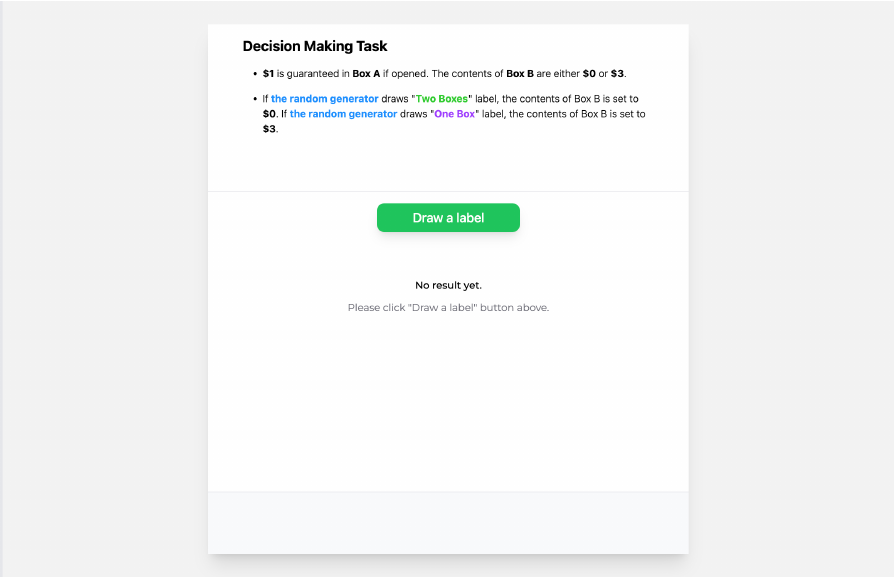}}
\end{center}
\begin{center}
    \fbox{\includegraphics[width=0.8\textwidth]{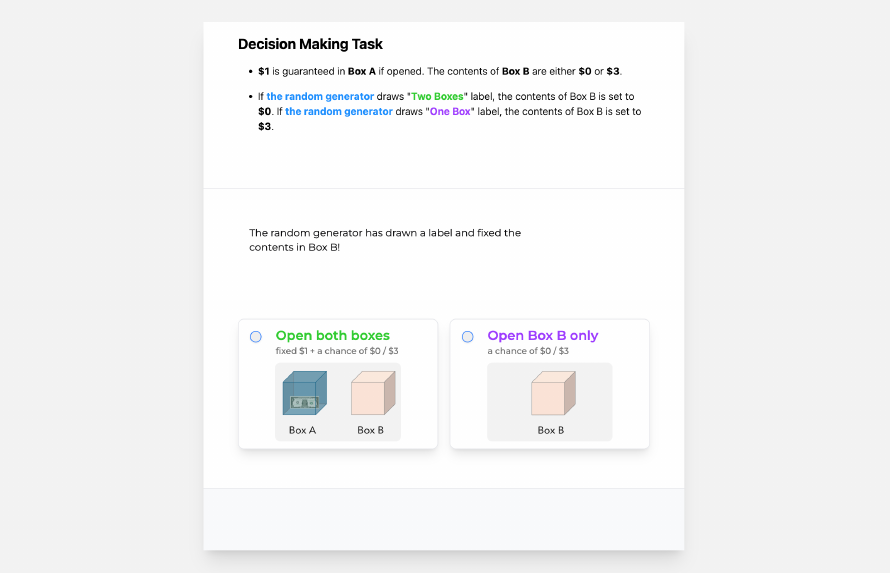}}
\end{center}
\begin{center}
    \fbox{\includegraphics[width=0.8\textwidth]{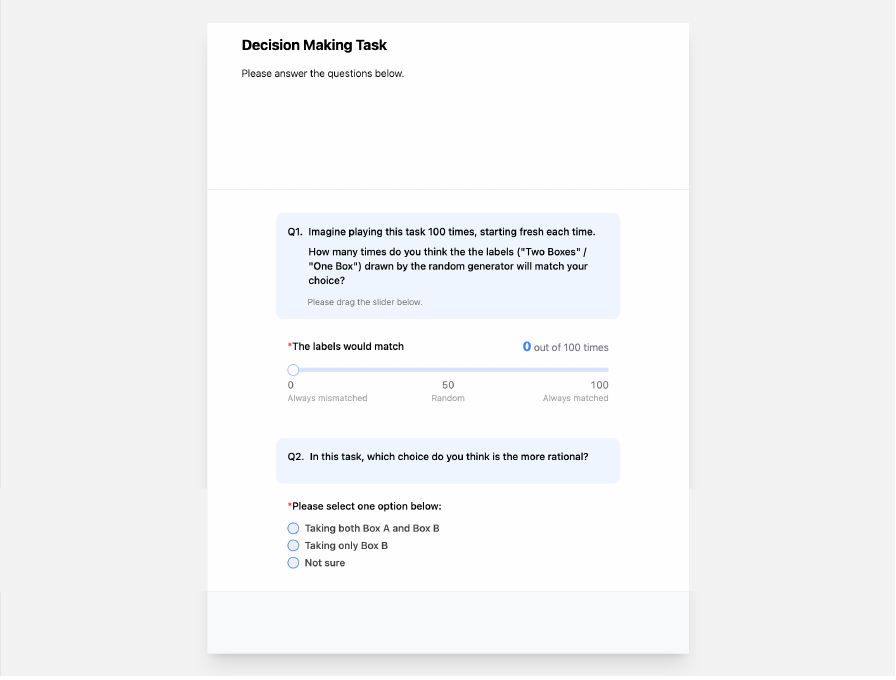}}
\end{center}
\begin{center}
    \fbox{\includegraphics[width=0.8\textwidth]{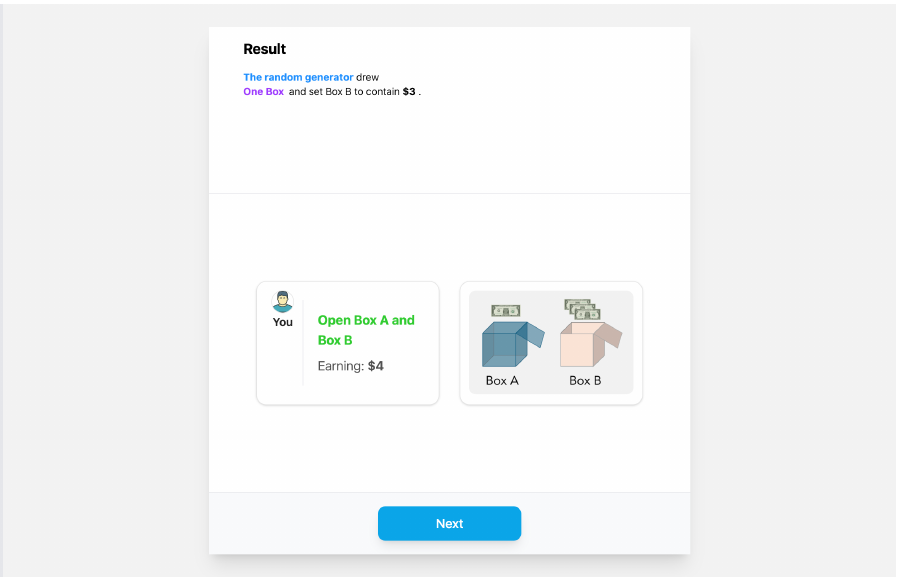}}
\end{center}
\newpage
These two pages are only for the interactive random condition.
\begin{center}
    \fbox{\includegraphics[width=0.8\textwidth]{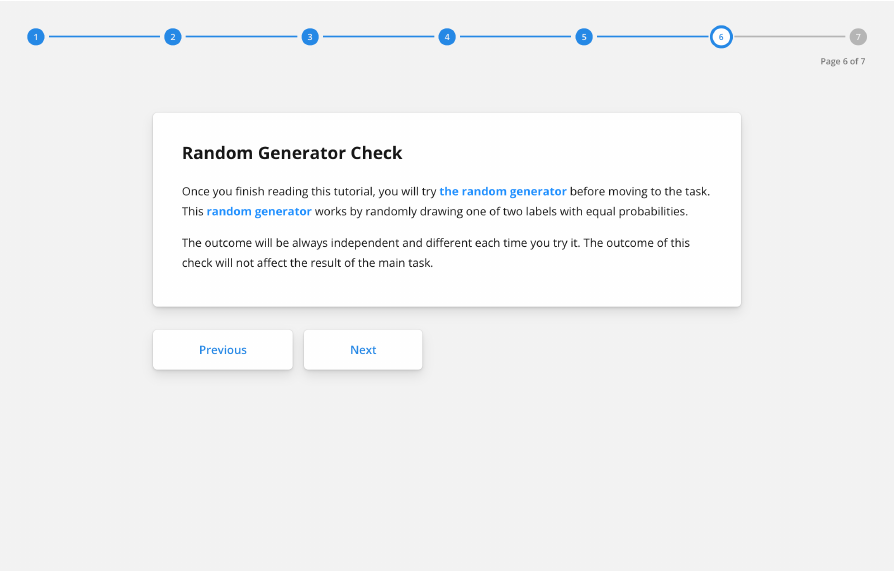}}
\end{center}
\begin{center}
    \fbox{\includegraphics[width=0.8\textwidth]{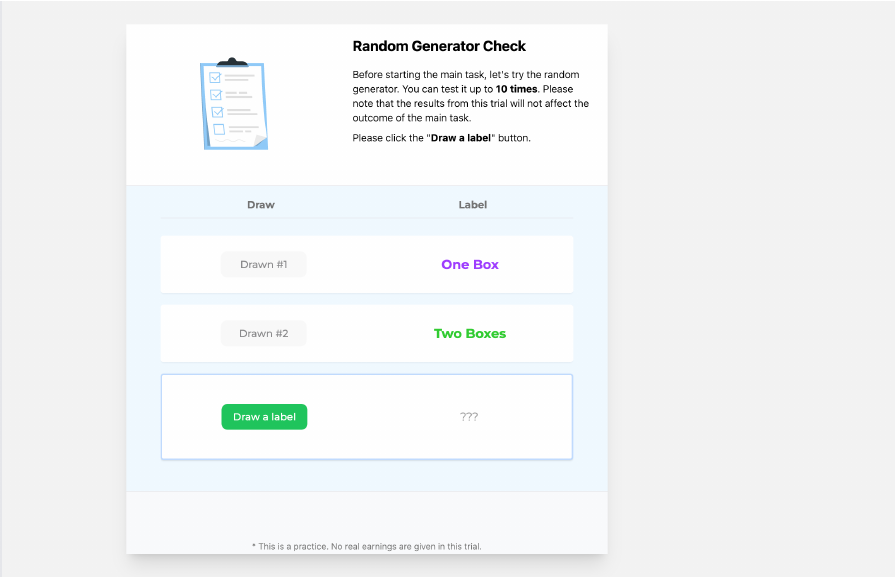}}
\end{center}
\newpage
These two pages are only for Study 4.
\begin{center}
    \fbox{\includegraphics[width=0.8\textwidth]{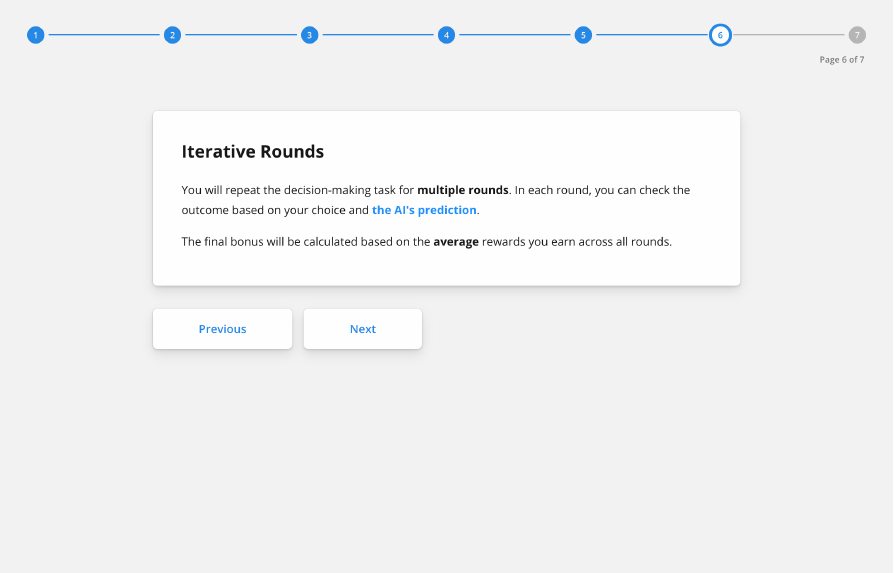}}
\end{center}
\begin{center}
    \fbox{\includegraphics[width=0.8\textwidth]{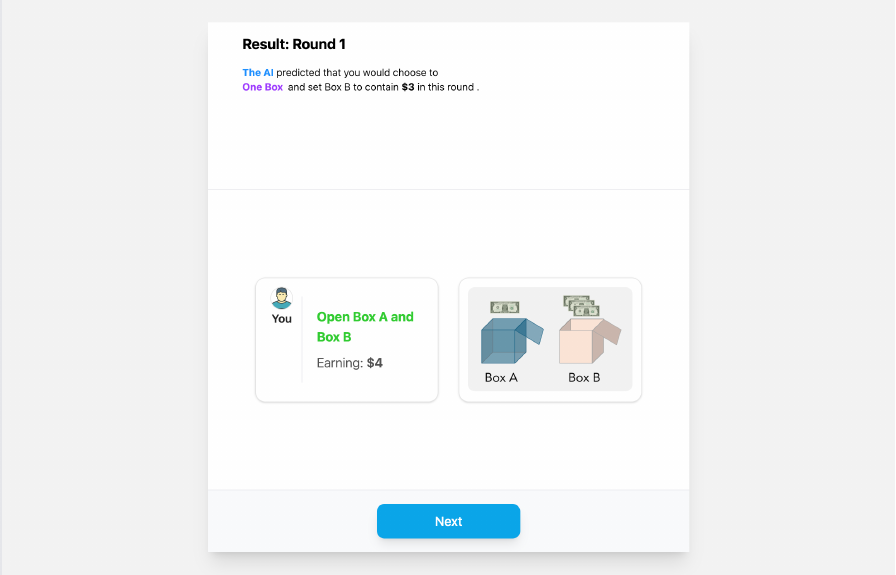}}
\end{center}
\newpage

\subsubsection*{Statistical analyses}
\paragraph*{A fixed-effect meta-analysis across Studies 1 and 2.}
To evaluate the overall effect of the AI condition on participants' decisions across Studies 1 and 2, we conducted a fixed-effect meta-analysis. First, we independently fitted logistic regression models for both studies. The dependent variable was the binary choice in the task (1 = one-boxing, 0 = two-boxing), and the independent variable was the identity of the predictor (1 = AI, 0 = random).

Next, we pooled the extracted log-odds coefficients ($\beta$) and their standard errors ($SE$) from both models using a fixed-effects meta-analysis with inverse-variance weighting. The pooled standard error ($SE_{meta}$), $Z$-score, and the corresponding $p$-value were calculated to assess the overall statistical significance. The pooled effect size was then converted into an odds ratio (OR) with a 95\% confidence interval (CI).

Finally, to examine the consistency of the AI effect across the two studies, we assessed between-study heterogeneity using Cochran's $Q$ statistic and the $I^2$ index.

\paragraph*{Competitive ratio of realized earnings to theory boundaries under AI regimes.}
To evaluate the observed shift toward one-boxing reduced realized earnings in Studies 1 and 2, we calculated competitive ratio of realized to theoretical earnings under AI regimes. The lower bound corresponds to a prediction regime in which the AI always predicts one-boxing (yielding payoffs of $3$ or $4$), whereas the upper bound corresponds to a regime in which the AI always predicts two-boxing (yielding payoffs of $0$ or $1$).

\subsubsection*{Computational modeling of decision-making}

\paragraph*{Model definition}
To examine how perceived predictiveness and internal coherence contribute to participants' actions in Study 2, we introduced a computational model of the decision-making process grounded in subjective expected utility theory \cite{gilboa2009theory}. 

Let $A$ denote the participant's choice (one-boxing or two-boxing), $P$ denote the system's prediction of $A$, and $AA$ denote the participant's anticipated action. 
Perceived predictiveness is defined as $p = \Pr(P = AA)$ and internal coherence as $q = \Pr(A = AA)$. 
In this formulation, $p$ represents the perceived probability that the system's prediction matches the participant's anticipated action, and $q$ represents the extent to which participants expect their eventual action to be consistent with their anticipated action (Fig.~\ref{fig:mechanism}A). 

Box A (transparent) always yields payoff $S$, whereas Box B (opaque) yields either $0$ or $L$ depending on the system's prediction $P$, with $0 < S < L$ (Fig. \ref{fig:main}A).
Participants are assumed to evaluate actions using subjective expected utility. 
The expected utilities of one-boxing and two-boxing, denoted by $EU[\text{1box}]$ and $EU[\text{2box}]$, depend on both perceived predictiveness and internal coherence.


If participant $i$ assumes no internal coherence  between anticipated and realized action (i.e., $q_i=0$), then the chosen action provides no information about the participant’s anticipated action. 
Under this assumption, the expected payoff from Box B is the same for one-boxing and two-boxing and depends only on the perceived predictiveness parameter $p_i$. 
The expected utilities become

\begin{equation}
    EU[\text{1box}\mid q_i = 0] = p_i L
\end{equation}

\begin{equation}
    EU[\text{2box}\mid q_i = 0] = S + p_i L
\end{equation}

\begin{equation} \label{eu_independent}
    \Delta EU_{\text{independent}}
    = EU[\text{1box}] - EU[\text{2box}]
    = -S .
\end{equation}
Thus, one-boxing is never preferred.

If participant $i$ assumes that anticipated action and realized action are perfectly coherent (i.e., $q_i=1$), then the chosen action fully reveals the participant's anticipated action. 
Under this assumption, the expected content of Box B depends on both perceived predictiveness $p_i$ and the participant's choice. 
The expected utilities become

\begin{equation}
    EU[\text{1box}\mid q_i=1] = p_i L
\end{equation}

\begin{equation}
    EU[\text{2box}\mid q_i=1] = S + (1 - p_i)L
\end{equation}

\begin{equation} \label{eu_coherent}
    \Delta EU_{\text{coherent}}(p_i)
    = 2p_iL - L - S .
\end{equation}
Thus, one-boxing is preferred whenever $p_i > (L+S)/(2L)$.

Under these assumptions, we evaluated four candidate models of decision-making: (i) the random choice model, (ii)  the causal reasoning model, (iii) the evidential reasoning model, and (iv) the mixture reasoning model. 
Each model specifies different assumptions about perceived predictiveness and internal coherence, as described below.

\paragraph*{Random choice model}
As a baseline, we assume that participants choose randomly between the two options with equal probability, independent of prediction or anticipation:
\begin{equation}
    \Pr(\text{1box}_i) = 0.5 .
\end{equation}

\paragraph*{Causal reasoning model}
In the causal reasoning model, we assume that participants ignore any coherence between anticipated action and realized action (i.e., $q_i=0$). 
Decisions are therefore based only on the expected utility difference in Eq.~\ref{eu_independent}. The probability of choosing one-boxing is

\begin{equation}
    \Pr(\text{1box}_i) = \sigma(\tau \cdot \Delta EU_{\text{independent}})
\end{equation}

where $\sigma(x) = \frac{1}{1 + e^{-x}}$ is a logistic function and $\tau > 0$ is an inverse temperature parameter capturing sensitivity to the expected utility difference.

\paragraph*{Evidential reasoning model}
In the evidential reasoning model, we assume that participants perceive perfect coherence between their anticipated and realized actions (i.e., $q_i=1$). 
Decisions are therefore based on the perceived predictiveness $p_i$ and the expected utility difference in Eq.~\ref{eu_coherent}. The probability of choosing one-boxing is

\begin{equation}
    \Pr(\text{1box}_i) = \sigma(\tau \cdot \Delta EU_{\text{coherent}}(p_i)).
\end{equation}

\paragraph*{Mixture reasoning model}
In the mixture reasoning model, participants are assumed to evaluate decisions using a mixture of causal and evidential reasoning. 
Let $q_i \in [0,1]$ denote participant $i$’s degree of internal coherence. Greater internal coherence increases the extent to which participants evaluate the decision through evidential rather than causal reasoning, conditional on their perceived predictiveness $p_i$.

The resulting choice probability is given by

\begin{equation}
    \Pr(\text{1box}_i) =
    q_i \, \sigma(\tau \cdot \Delta EU_{\text{coherent}}(p_i))
    + (1 - q_i) \, \sigma(\tau \cdot \Delta EU_{\text{independent}}).
\end{equation}

\paragraph*{Parameter estimation}
The observed choice $y_i \in \{0,1\}$ (where $1$ denotes the one-boxing choice) for participant $i$ is modeled as a Bernoulli trial drawn from the predicted probability:

\begin{equation}
    y_i \sim \text{Bernoulli}(\Pr(\text{1box}_i)).
\end{equation}


Given the between-subject design, in which each participant made a one-shot binary choice, we adopted a Bayesian approach to stabilize parameter estimation. 
Prior to estimation, perceived predictiveness, originally measured on a scale from 0 to 100, was rescaled to the unit interval $[0,1]$.

In Study 2, perceived predictiveness was measured directly for each participant via their self-reported survey and therefore entered the model as an individual-level observed variable $p_i$.
Internal coherence, however, could not be estimated reliably at the individual level because each participant made only a single binary decision.
We therefore estimated a condition-specific coherence parameter $q_c$, shared by participants within each experimental condition ($c \in \{\text{AI}, \text{random}\}$). 

To regularize estimation while remaining weakly informative, we assigned weakly informative priors to all free parameters. 
\begin{equation}
    \tau \sim \text{HalfNormal}(2)
\end{equation}
\begin{equation}
    q_{c} \sim \text{Beta}(2,2), \quad c \in \{\text{AI}, \text{random}\}.
\end{equation}

Posterior distributions were estimated using the No-U-Turn Sampler (NUTS), a variant of Hamiltonian Monte Carlo, implemented in PyMC (v5.27.0). We ran four parallel chains with 2,000 iterations each, discarding the first 1,000 iterations of each chain as warmup, yielding 4,000 posterior samples. Convergence was assessed using the Gelman–Rubin statistic ($\hat{R}$), and all parameters satisfied $\hat{R} < 1.01$.

\paragraph*{Modeling results}
First, we compared the goodness-of-fit of the four candidate models using LOO, WAIC, and WBIC. 
The mixture reasoning model provided the best fit across all metrics (Extended Data Table~\ref{tab:study02_model_selection}). 
This result suggests that variation in perceived predictiveness alone cannot fully account for the observed choice patterns, and that incorporating an internal coherence component improves the model's explanatory power.

To further examine how prediction framing influenced participants’ behavior, we inspected the posterior distributions of the parameters in the winning model. 
The inverse temperature parameter ($\tau$) was estimated at 0.263 (95\% HDI: [0.216, 0.315]), indicating that participants' choices were systematically related to the expected utility differences implied by the model. 

We then compared the estimated internal-coherence parameters for the AI condition ($q_{AI}$) and the random control condition ($q_{random}$).
In the model, greater internal coherence corresponds to a higher probability of evaluating the task through evidential rather than causal reasoning.
The posterior mean of $q_{AI}$ was 0.682 (95\% HDI: [0.425, 0.927]), whereas the posterior mean of $q_{random}$ was lower at 0.443 (95\% HDI: [0.076, 0.836]). 
However, the posterior probability that $q_{AI} > q_{random}$ was 82.6\%, which does not reach a conventional evidential threshold (e.g., 95\%).

Together, these results indicate that a model combining perceived predictiveness with a mixture of causal and evidential reasoning provides the best account of the observed behavior.
Models assuming solely causal reasoning provided a substantially poorer fit than models allowing for evidential reasoning.
These findings are consistent with the interpretation that participants used a mixture of causal and evidential reasoning when evaluating AI predictions, with evidential reasoning arising from both perceived predictiveness and internal coherence.
They further suggest that AI framing may increase the tendency to evaluate decisions through evidential rather than causal reasoning, although the present data do not provide decisive statistical evidence for this difference.

\end{document}